\documentclass[fleqn,usenatbib,useAMS]{mnras}

\usepackage{svg}
\usepackage{graphicx}	% Including figure files
\usepackage{amsmath}	% Advanced maths commands
\usepackage{amssymb}	% Extra maths symbols
\usepackage[super]{nth}

\usepackage[T1]{fontenc}
\usepackage{ae,aecompl}

\usepackage{newtxtext,newtxmath}

\usepackage{xcolor}
\usepackage[normalem]{ulem}

\DeclareRobustCommand{\VAN}[3]{#2}
\let\VANthebibliography\thebibliography
\def\thebibliography{\DeclareRobustCommand{\VAN}[3]{##3}\VANthebibliography}

\title[SILCC VI - Thermal \& non-thermal ISM]{SILCC VI - Multi-phase ISM structure, stellar clustering, and outflows with supernovae, stellar winds, ionising radiation and cosmic rays}

\author[T.-E. Rathjen et al.]{
Tim-Eric Rathjen,$^{1}$\thanks{E-mail: rathjen@mpa-garching.mpg.de}
Thorsten Naab,$^{1}$
Philipp Girichidis,$^{2}$
Stefanie Walch,$^{3}$
Richard W\"unsch,$^{4}$
\newauthor
Franti\u{s}ek Dinnbier,$^{5}$
Daniel Seifried,$^{3}$
Ralf S. Klessen$^{6, 7}$
and Simon C. O. Glover$^{6}$
\\
$^{1}$Max Planck Institute for Astrophysics, Karl-Schwarzschild-Str. 1, 85748 Garching, Germany\\
$^{2}$Leibniz Institute for Astrophysics, An der Sternwarte 16, 14482 Potsdam, Germany\\
$^{3}$I. Physikalisches Institut, Universit\"at zu K\"oln, Z\"ulpicher Str. 77, 50937 K\"oln, Germany\\
$^{4}$Astronomical Institute of the Czech Academy of Sciences, Bo\v{c}n\'{i} II 1401/1, 141 00 Praha 4, Czech Republic\\
$^{5}$Charles University in Prague, Faculty of Mathematics and Physics, Astronomical Institute, V Hole\u{s}ovi\u{c}k\'{a}ch 2, 180 00 Praha 8, Czech Republic\\
$^{6}$Universit\"at Heidelberg, Zentrum f\"ur Astronomie, Institut f\"ur Theoretische Astrophysik, Albert-Ueberle-Str. 2, 69120 Heidelberg, Germany\\
$^{7}$Universit\"at Heidelberg, Interdisziplin\"ares Zentrum f\"ur Wissenschaftliches Rechnen, Im Neuenheimer Feld 205, 69120 Heidelberg, Germany
}

\date{Accepted 2021 March 25. Received 2021 March 19; in original form 2021 January 8}

\pubyear{2020}

\begin{document}
\label{firstpage}
\pagerange{\pageref{firstpage}--\pageref{lastpage}}
\maketitle
 
\begin{abstract}
We present simulations of the multi-phase interstellar medium (ISM) at solar neighbourhood conditions including thermal and non-thermal ISM processes, star cluster formation, and feedback from massive stars: stellar winds, hydrogen ionising radiation computed with the novel \textsc{TreeRay} radiative transfer method, supernovae (SN), and the injection of cosmic rays (CR). N-body dynamics is computed with a \nth{4}-order Hermite integrator. We systematically investigate the impact of stellar feedback on the self-gravitating ISM with magnetic fields, CR advection and diffusion and non-equilibrium chemical evolution. SN-only feedback results in strongly clustered star formation with very high star cluster masses, a bi-modal distribution of the ambient SN densities, and low volume-filling factors (VFF) of warm gas, typically inconsistent with local conditions. Early radiative feedback prevents an initial starburst, reduces star cluster masses and outflow rates. Furthermore, star formation rate surface densities of $\Sigma_{\dot{M}_\star} = 1.4-5.9 \times 10^{-3}$ $\mathrm{M}_\odot\,\mathrm{yr}^{-1}\,\mathrm{kpc}^{-2}$, VFF$_\mathrm{warm} = 60-80$ per cent as well as thermal, kinetic, magnetic, and cosmic ray energy densities of the model including all feedback mechanisms agree well with observational constraints. On the short, 100 Myr, timescales investigated here, CRs only have a moderate impact on star formation and the multi-phase gas structure and result in cooler outflows, if present. Our models indicate that at low gas surface densities SN-only feedback only captures some characteristics of the star-forming ISM and outflows/inflows relevant for regulating star formation. Instead, star formation is regulated on star cluster scales by radiation and winds from massive stars in clusters, whose peak masses agree with solar neighbourhood estimates.
\end{abstract}


\section{Introduction}\label{sec:intro}

The interstellar medium (ISM) is traditionally defined as \textit{everything} in-between the stars in galaxies, accounting for the non-stellar and non-relativistic baryonic matter, radiation, magnetic fields, and cosmic rays (CR) in galactic discs. In the ISM, star formation takes place, galactic outflows are launched, and galaxies grow in size and mass. Through an inflow/outflow interface, the ISM smoothly transitions to the circumgalactic medium (CGM) occupying the more spherical galactic halo. The CGM can contain significant fractions of the total baryonic mass which is gravitationally bound to galaxies but shows no evidence for star formation \citep{Tumlinson2017}.

The ISM is of a multi-phase nature with ionised, neutral, and molecular gas as well as dust \citep{Draine2011}. The cold and warm components are believed to be in pressure equilibrium \citep{Wolfire2003, Cox2005a}. An additional meta-stable hot phase exists with gas temperatures exceeding $T = 10^5$ K \citep{Cox1974a, McKee1977, Ferriere2001, Klessen2016} generated predominately by supernova (SN) explosions.

Molecular gas is typically found in structured and compact molecular clouds, where all new stars in galaxies are born \citep{McKee2007, KennicuttJr.2012}. Those molecular clouds can be formed by cooling and gravitational collapse of the magnetised gas in dust shielded regions \citep{KennicuttJr.2012, Ibanez-Mejia2017} or by sweeping up gas and supersonic compression from multiple SN explosions \citep{Inutsuka2015, Seifried2017}. 
Most of the volume in the ISM, however, is occupied by neutral and ionised gas. Interstellar radiation from stars or gas cooling processes is also part of the ISM \citep{Ferriere2001}. 
Additionally, magnetic fields and CRs – typically protons at relativistic speeds – are energetically equally important non-thermal components \citep{Draine2011, Heitsch2009, Crutcher2012} and might play a vital role in the evolution of galaxies \citep[see e.g.][for an overview]{Naab2017}.

Feedback from massive O and B stars has the strongest impact on the environment by injecting radiation, momentum, and energy into the ISM \citep{MacLow2004, Krumholz2014, Haid2018}. Massive stars form in clusters \citep{Lada2003} and create HII regions by ionising and heating their surroundings with UV radiation \citep[see e.g.][]{Spitzer1978, Whitworth1979, Dale2005, Dale2012, Walch2012a, Walch2013a, Dale2014, Geen2015, Haid2018, Haid2019, Kim2020}. Additionally, stellar winds partly disperse their parental clouds \citep[see e.g.][]{Castor1975, Weaver1977, Wunsch2008, Wunsch2011, Toala2011, Dale2012, Rogers2013a, Mackey2015, Haid2018}. At the end of a massive stars lifetime, SNe drive strong shocks into the ISM by generating hot ionised gas in expanding super-bubbles \citep[see e.g.][]{MacLow1988, MacLow1989, Gatto2015, Kim2015, Martizzi2015, Walch2015a, Walch2015, Haid2016}. CRs generated in these shocks interact with the magnetic field and generate an additional pressure component whose gradient can drive gas out of the ISM \citep{Dorfi2012, Simpson2016, Girichidis2016, Girichidis2018}. Local observations suggest that CRs are accelerated by diffusive shock acceleration in SN remnants \citep{Bell1978, Blandford1978} with an efficiency of $\sim10$ per cent \citep{Helder2012, Ackermann2013}. CRs have energy densities $u_{\mathrm{cr}} \approx 1.4\,\mathrm{eV~cm}^{-3}$, comparable to the thermal, turbulent, and magnetic energy densities \citep{Draine2011}. The impact of this relativistic component has only recently been investigated in numerical ISM and galaxy formation studies (e.g. \citealt{Hanasz2013, Booth2013, Salem2014, Girichidis2016, Pakmor2016, Simpson2016, Girichidis2018}).

The most dramatic single events, however, are the blast waves generated by SNe. They have a considerable dynamical impact on the ISM \citep{McKee1977, MacLow2004}. The SN impact can be stronger if they explode in low-density environments \citep{Creasey2013, Martizzi2015, Gatto2015, Iffrig2015, Kim2015, Walch2015a, Fielding2017}, otherwise, their injected energy is typically radiated away without strong coupling to the ambient gas \citep{Walch2015a, Naab2017}, even to the point that no Sedov-Taylor stage is developed \citep{Jimenez2019}. The non-linear interaction of clustered star formation, thermal and non-thermal feedback processes with the highly structured multi-phase ISM can be best investigated with numerical simulations. 

Idealised stratified galactic disc simulations have followed the evolution of the multi-phase ISM and outflows driven by SNe with fixed rates inferred from observations \citep[e.g.][]{DeAvillez2005, Joung2006, Hill2012, Gent2013, Walch2015, Girichidis2016, Li2015, Li2017}. Such approaches, however, do not allow for a self-consistent study of the evolution of the star-forming ISM. 
Current approaches allow for the modelling of star formation via sink particles \citep{Federrath2010, Gatto2017, Peters2017, Iffrig2017, Kim2017, Kim2018} and include feedback from massive stars by SNe \citep[e.g.][]{Gatto2017, Kim2017}, SNe and stellar winds \citep{Gatto2017}, SNe and radiation either directly \citep[e.g.][]{Butler2017} or in post-processing \citep{Kado-Fong2020}, or SNe, stellar winds and radiation combined \citep{Peters2017}. ISM studies on the impact of CRs have, so far, no self-consistent star formation included.

Within the \textsc{SILCC} project\footnote{\url{https://hera.ph1.uni-koeln.de/~silcc/}} and related publications, the governing processes setting the ISM structure have been studied in idealised experiments. We have subsequently included feedback processes from SNe \citep{Walch2015a, Girichidis2016a}, stellar winds \citep{Gatto2017}, ionising radiation \citep{Peters2017} and magnetic fields \citep{Pardi2017, Girichidis2018a}. There are strong indications that radiation impacts and reduces the SFR and qualitatively changes the ISM structure. Therefore, this process has to be taken into account in studies aiming at creating a realistic model of the multi-phase ISM.

In this paper, we combine all of of the aforementioned processes using the novel radiation transfer method \textsc{TreeRay} and a novel implementation of a \nth{4}-order Hermite integrator for computing the N-body dynamics of the stellar cluster sink particles. We also investigate additional injection of CR in SNe and their propagation. We thereby present a set of self-consistent, parsec-scale, stratified disc MHD simulations of the solar neighbourhood at increasing physical complexity and realism. 

This paper is structured as follows: In Sec. \ref{sec:methods}, we introduce the physical modules and explain the simulation setup. In Sec. \ref{sec:morphology}, we give an overview of the global evolution and morphology of the runs. The complexity of stellar feedback is analysed in Sec. \ref{sec:feeback}, with a focus on the star formation properties in Sec. \ref{sec:sfr}, star cluster formation in Sec. \ref{sec:cluster}, and SN impact in Sec. \ref{sec:sn}. In Sec. \ref{sec:structure}, we investigate the ISM structure and their mass- and volume-filling factors. In Sec. \ref{sec:outflow}, we study the impact of stellar feedback on galactic outflows. A discussion about our work in the context of other studies and possible caveats of our models are given in Sec. \ref{sec:discussion}, and the paper is summarised and concluded in Sec. \ref{sec:summary}. We briefly present the phase structure of the outflow in Appendix \ref{sec:phase_outflow}, the energy injection of the different stellar feedback mechanisms in Appendix \ref{sec:inj} and a short discussion about a possible different realisation for SN injection in Appendix \ref{sec:flat}.

\begin{figure*}
	\centering
	\includegraphics[width=1\linewidth]{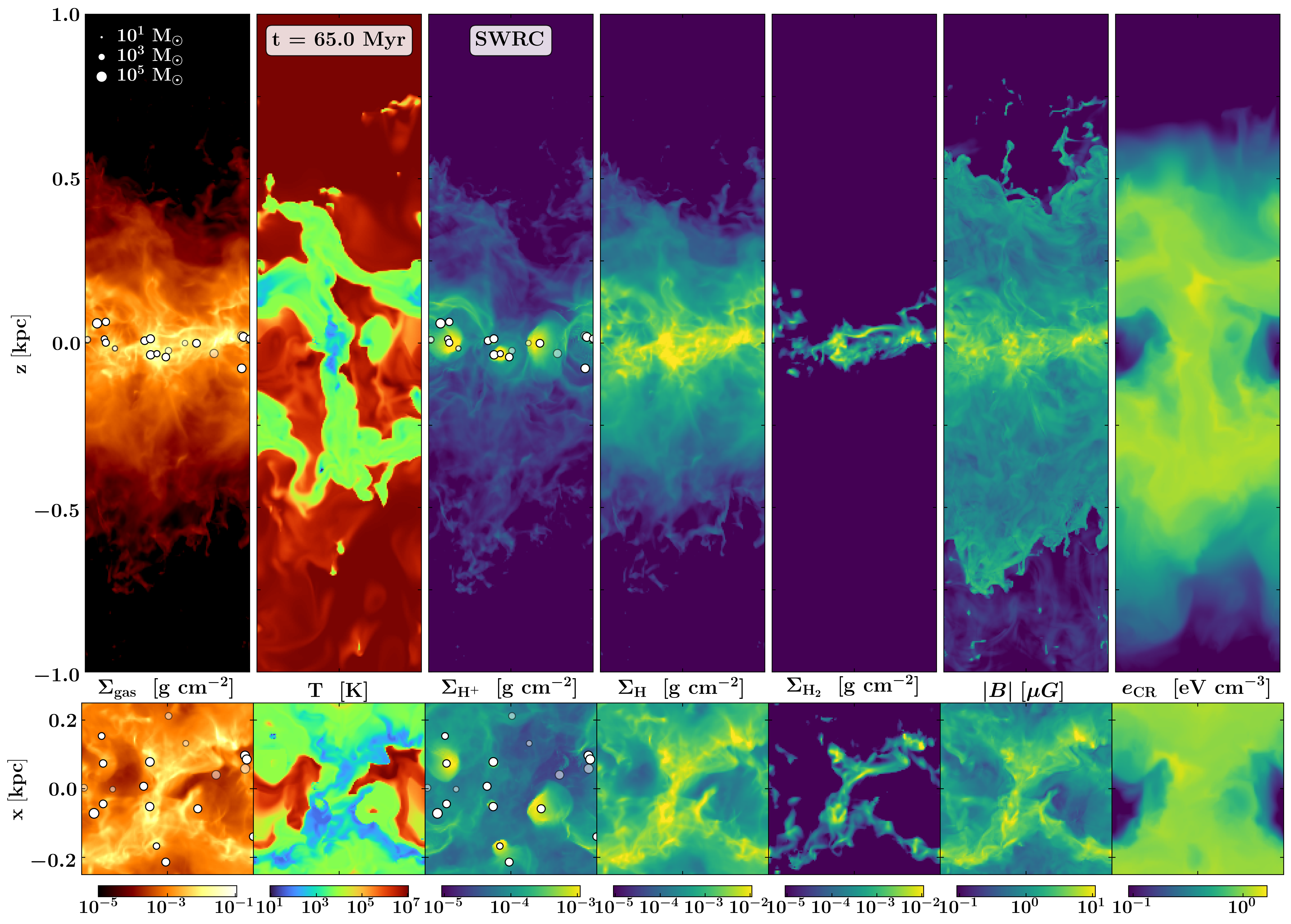}
	\caption{Overview of the \textit{SWRC} (see Table \ref{tab:runs}) run, including supernovae, stellar winds, UV radiation and cosmic rays, at $t = 65$ Myr. Shown are the edge-on (top row) and face-on (bottom row) views of the total gas, ionised-, atomic-, and molecular hydrogen column densities. Individual HII regions (\nth{3} panel) from active star clusters are visible. We also show the density-weighted column of the magnetic field strength (\nth{6} panel) and slices through the centre of the simulation box with temperature (\nth{2} panel) and CR energy density (\nth{7} panel). The star-forming galactic ISM is concentrated around the mid-plane. White circles in the \nth{1} and \nth{3} panel indicate star clusters with different masses. Translucent symbols indicate \textit{old} star clusters with no active massive stars in them. Stellar feedback generates a highly structured and turbulent multi-phase ISM with all its major thermal and non-thermal components. \label{fig:ov}}
\end{figure*}

\section{Numerical methods and simulation setup}\label{sec:methods}

Stratified disc patches are simulated using the MPI parallel, 3D adaptive mesh refinement magneto-hydrodynamics (MHD) code \textsc{FLASH} \citep{Fryxell2000, Dubey2008, Dubey2009}. Our setup follows the general \textsc{SILCC} framework \citep{Walch2015a, Girichidis2016a, Gatto2017, Peters2017, Girichidis2018a} with the inclusion of the radiative transfer solver \textsc{TreeRay} \citep[][W\"unsch et al., submitted]{Haid2019}, \nth{4}-order Hermite integrator for sink particle N-body dynamics \cite{Dinnbier2020}, and anisotropic CR transport as in \citet{Girichidis2016, Girichidis2018}. The MHD equations are solved with a modified, directionally split, three-wave Bouchut scheme (HLLR3) for ideal MHD, suitable for flows of high Mach number \citep{Bouchut2007, Bouchut2010, Klingenberg2007, Waagan2011}.
Self-gravity is accounted for by solving the Poisson equation via an Oct-tree based method \citep{Barnes1986, Wunsch2018}. An external potential is also included to model the gravitational impact of the pre-existing stellar disc and the contribution of a dark matter halo. 

We explicitly follow the non-equilibrium time-dependent chemical evolution of H, H$^+$, H$_2$, C$^+$, CO \citep{ Nelson1997, Glover2007} and account for photoelectric heating and radiative cooling, assuming a constant far ultraviolet (FUV) interstellar radiation field (ISRF) with $G_0 = 1.7$ \citep{Draine1978} and a constant dust-to-gas ratio of 1 per cent. The local optical depth of gas and dust and hence their shielding and self-shielding is calculated with the \textsc{TreeCol} algorithm \citep{Clark2012, Wunsch2018}. We refer the reader to \citet{Walch2015a} for a detailed description of the chemical network and the shielding processes included.

Star formation is modelled with accreting Lagrangian sink particles, which represent star clusters \citep[see][]{Gatto2017}. For each cluster sink, we explicitly follow the evolution of massive stars in a mass range $m_\star = 9-120$ M$_\odot$. For every 120 M$_\odot$ of accreted gas, we form one new massive star sampled from a Salpeter IMF \citep{Salpeter1955}. Accretion and formation of the sink particles are described by \citet{Federrath2010}, with an accretion radius of $r_\mathrm{accr} = 3 \times \Delta x$ ($\sim11.7\,\mathrm{pc}$ at the highest refinement level with $\Delta x \approx 3.9\,\mathrm{pc}$) and a particle threshold density of $n_\mathrm{sink} \approx 10^3\,\mathrm{cm}^{-3}$. Furthermore, the gas within $r_\mathrm{accr}$ has to be in a converging flow, gravitationally bound, Jeans unstable, and in a local gravitational potential minimum to form or be accreted by a sink particle.
The accretion radius depends on the grid resolution and chosen to be as small as possible without creating grid artefacts \citep[see e.g.][]{Federrath2010, Hennebelle2014, Gatto2017, Peters2017}. If all accretion criteria are fulfilled, the gas that is above the threshold density within the accretion radius is added to the sink particle. It is important to note that not all the gas of the respective cells is accreted by the sink particle but only the difference between its actual density and the threshold density. Otherwise, the sink particles would create holes in the density structure of the ambient medium and potentially trigger runaway collapse. Therefore, there is no inherent minimum star cluster sink particle mass. Furthermore, the total gas density of cells within a sink particle's accretion radius is not necessarily at the threshold density of $n_\mathrm{sink} \approx 10^3\,\mathrm{cm}^{-3}$. We note that many cells within a sink particle's accretion radius are below that density.
The trajectories of the cluster sinks are computed with an \nth{4}-order Hermite predictor-corrector integration scheme \citep[see][for details]{Dinnbier2020}. The sink particles are coupled to the Oct-tree which makes the calculations of their interaction with the gas efficient for a large number of particles.

Type II supernovae (SNe) are realised by injecting ${E_\mathrm{sn} = 10^{51}}\,\mathrm{erg}$ as thermal energy into a spherical region with fixed radius ${r_\mathrm{inj} = 3 \times \Delta x}$ ($\sim11.7$ pc) around the sink particle in which a massive star explodes. We evenly distribute the ejecta mass in the same region and keep the density fluctuations in the injection region intact, i.e. we do not \textit{by hand} flatten the density to an average value. 

The SN remnant radius at the end of the Sedov-Taylor-phase \citep{Blondin1998} is
\begin{equation}
    R_\mathrm{ST} = 19.1 \left(\frac{E_\mathrm{SN}}{10^{51} \mathrm{erg}}\right)^{5/17}\left(\frac{\overline{n}}{\mathrm{cm}^{-3}}\right)^{-7/17} \mathrm{pc}.
\end{equation} 
To resolve this radius with at least 3 grid cells, the ambient density of a SN explosion site must not exceed $n_\mathrm{ambient} = 3.3$ cm$^{-3}$. If the average ambient density is above this threshold, we switch to momentum injection and deposit the expected radial blast wave momentum at the beginning of the momentum conserving snowplough phase into the injection region. Furthermore, we set the temperature of the injection region to $T = 10^4 \mathrm{K}$ as described in \citet{Gatto2017}. 

To account for stellar winds, we inject mass and radial momentum of each massive star in a cluster sink using mass loss rates form the Geneva stellar evolution tracks from the zero-age main sequence to the Wolf-Rayet phase \citep{Ekstrom2012}. The terminal wind velocities are estimated according to \citet{Puls2008}. The mass of the wind is evenly distributed in the injection region and the wind is assumed to be spherically symmetric. Chemical abundances in the injection region are kept unchanged. All details are discussed in \citet{Gatto2015}.

The propagation of ionising UV photons from massive stars is handled with the novel \textsc{TreeRay} algorithm (W\"unsch et. al. submitted) which has been benchmarked and applied in \citet{Bisbas2015} and \citet{Haid2018, Haid2019}. It is a backwards ray-tracing scheme which uses the Oct-tree structure from the gravity- and diffuse radiation solver described in \citet{Wunsch2018}. The method couples self-consistently to the chemistry using one energy bin for photons with energy ${h\nu \geq 13.6\,\mathrm{eV}}$. Heating by the UV is calculated using the mean excess photon energy as described in \citet{Haid2019}. First, the gas with its emission and absorption coefficients and the emitting sources are mapped onto the Oct-tree. Rays are then cast from each target cell via the \textsc{HEALPix} algorithm \citep{Gorski2005}, which distributes the rays uniformly over the surface of a unit sphere. Then, the 1D radiative transport equation is solved along each ray accounting for the radiation passing through the calculated ray from other directions. Finally, the whole process is repeated until the radiation density converges everywhere. The great advantage of this approach is that the cost of computation does not depend on the number of sources and hence multiple radiating star clusters can be handled effectively.
The UV photons coming from the massive stars are injected within the star cluster sink particle's accretion radius. However, the photons will already be locally absorbed and re-processed by the gas in the cells in which they get injected, and then propagated by the aforementioned mechanism. The photons are not launched from the surface of the sink particles but their centre. The absorption of the UV photons within the sink particle's radius is treated by the radiation transfer module in the same way as in the other cells along the UV photon's propagation. Nonetheless, local porosity and clumps cannot be resolved on scales below the cell size of $\Delta x = 4\,\mathrm{pc}$. The UV photon escape fraction from compact and ultra-compact HII regions (cHII) is an uncertainty in our models for resolution reasons. Still, the lifetime of cHII regions is of order $\sim0.3\,\mathrm{Myr}$ \citep{Mottram2011}, less than $\sim10$ per cent of the lifetime of the massive stars powering the HII regions. We do not expect this uncertainty to have a large impact on our results.
Photoelectric heating and photo-dissociation of H$_2$ is not treated by the radiative transfer module but is instead included through the assumed-uniform ISRF, which is attenuated at high column densities. The current prescription simplifies the role of dust, which can either decrease UV ionisation by attenuation or on the other hand enhance the UV escape fraction by excavating the centre of HII regions via radiation pressure, which is not included in our models. We justify the omission of radiation pressure by noting that radiation pressure is only expected to play an important role in molecular cloud dispersal on scales smaller than those that we resolve \citep{Olivier2020} and for star clusters more massive than the ones that form in our simulations \citep{Rahner2017, Reissl2018}. For molecular cloud scales like in our models the UV radiation will quickly be absorbed and re-emitted in the thermal infrared, at which wavelength the clouds are optically thin.

CRs are treated as an additional non-thermal, relativistic fluid in the advection-diffusion approximation. They add another source term $Q_\mathrm{cr}$ to the MHD equations \citep{Girichidis2016, Girichidis2018}, including the injection of CRs by SNe with an efficiency of 10 per cent (i.e. $E_\mathrm{cr} = 10^{50}\,\mathrm{erg}$, \citealt{Helder2012, Ackermann2013}) as well as hadronic losses $\Lambda_\mathrm{hadronic}$ as described in \citet{Pfrommer2017} and \citet{Girichidis2020}. We assume a steady-state energy spectrum. For the CR diffusion tensor, we choose $K_\parallel = 10^{28}$ cm$^2$ s$^{-1}$ parallel to the magnetic field lines and $K_\perp = 10^{26}$ cm$^2$ s$^{-1}$ perpendicular to the magnetic field lines \citep{Strong2007, Nava2013}. With CRs added, the complete set of MHD equations reads
\begin{align}
\frac{\partial\rho}{\partial t} &+ \nabla \cdot (\rho\mathbf{v}) = 0\\
\frac{\partial\rho\mathbf{v}}{\partial t} &+ \nabla \cdot \left(\rho \mathbf{v}\mathbf{v}^\mathrm{T} - \frac{\mathbf{B}\mathbf{B}^\mathrm{T}}{4\pi}\right) + \nabla P_\mathrm{tot} = \rho\mathbf{g} + \dot{\mathbf{q}}_\mathrm{sn}\\
\frac{\partial e}{\partial t} &+ \nabla \cdot \left[\left(e + P_\mathrm{tot}\right)\mathbf{v} - \frac{\mathbf{B}\left(\mathbf{B} \cdot \mathbf{v}\right)}{4\pi}\right] \notag\\
&= \rho\mathbf{v} \cdot \mathbf{g} + \nabla \cdot \left(\mathsf{K}\nabla e_\mathrm{cr}\right) + \dot{u}_\mathrm{chem} + \dot{u}_\mathrm{sn} + Q_\mathrm{cr}\\
\frac{\partial\mathbf{B}}{\partial t} &- \nabla \times \left(\mathbf{v} \times \mathbf{B}\right) = 0\\
\frac{\partial e_\mathrm{cr}}{\partial t} &+ \nabla \cdot \left(e_\mathrm{cr}\mathbf{v}\right) = -P_\mathrm{cr}\nabla\cdot\mathbf{v}+\nabla\cdot\left(\mathsf{K}\nabla e_\mathrm{cr}\right) + Q_\mathrm{cr},
\end{align}
with the mass density $\rho$, the gas velocity $\mathbf{v}$, the magnetic field $\mathbf{B}$, the total pressure $P_\mathrm{tot} = P_\mathrm{thermal} + P_\mathrm{magnetic} + P_\mathrm{cr}$, the total energy density $e = \frac{\rho v^2}{2} + e_\mathrm{thermal} + e_\mathrm{cr} + \frac{B^2}{8\pi}$, the momentum input of unresolved SNe $\dot{\mathbf{q}}_\mathrm{sn}$, the thermal energy input from resolved SNe, $\dot{u}_\mathrm{sn}$, the changes in thermal energy due to heating and cooling, $\dot{u}_\mathrm{chem}$, the CR diffusion tensor, $\mathsf{K}$, and the CR energy source term, ${Q_\mathrm{cr} = Q_\mathrm{cr, injection} + \Lambda_\mathrm{hadronic}}$. The resulting effective adiabatic index is $\gamma_\mathrm{eff} = \frac{\gamma P_\mathrm{thermal} + \gamma_\mathrm{cr}P_\mathrm{cr}}{P_\mathrm{thermal} + P_\mathrm{cr}}$ with $\gamma = \frac{5}{3}$ and $\gamma_\mathrm{cr} = \frac{4}{3}$.

\subsection{Simulation parameters}\label{sec:setup}
We run a suite of six stratified box simulations. They all have a size of $0.5\,\mathrm{kpc} \times 0.5\,\mathrm{kpc} \times 4\,\mathrm{kpc}$ with periodic boundaries in $x$- and $y$- direction and strictly outflow boundary conditions in the $z$-direction, i.e. no material is allowed to flow back into the box.
Those boundary conditions do not allow for shearing flows and therefore the impact of galactic shear is not accounted for in this study. Possible ramifications of this omission are discussed in Sec. \ref{sec:discussion}.
Within ${z = \pm 1\,\mathrm{kpc}}$, we always adopt a resolution of ${\Delta x \approx 3.9\,\mathrm{pc}}$, whereas outside of this region we adopt a base resolution of $\Delta x \approx 7.8$ pc with the possibility to refine on the density gradient up to ${\Delta x \approx 3.9\,\mathrm{pc}}$. We set up the gas with a Gaussian distribution in $z$ and a scale height of 30 pc with a surface density of ${\Sigma_\mathrm{gas} = 10\,\mathrm{M}_\odot\,\mathrm{pc}^{-2}}$ and solar metallicity, mimicking solar neighbourhood conditions. The medium is magnetised with an initial magnetic field along the $x$-axis of the box and field strength of $B_x = 6$ $\mu$G. At the beginning of the simulation, the gas in the mid-plane is set to be in pressure equilibrium and purely atomic. We artificially drive large-scale turbulence for the first 10 Myr to introduce inhomogeneities and prevent the gas from collapsing into a thin sheet in the mid-plane. This is done by injecting kinetic energy on the largest scale corresponding to the box side-length $L_x = L_y = 0.5\,\mathrm{kpc}$ with a mix of 2:1 of solenoidal to compressive modes \citep{Schmidt2009, Konstandin2015} so that the gas stays at a constant root mean square velocity of ${v_\mathrm{rms} = 10\,\mathrm{km s}^{-1}}$ \citep{Eswaran1988}. For the external potential, we take an isothermal sheet \citep{Spitzer1942} with a stellar surface density of $\Sigma_\star = 30\,\mathrm{M}_\odot\,\mathrm{pc}^{-2}$ and a vertical scale height $z_\mathrm{d} = 300$ pc for the stars. For the dark matter, we assume an NFW profile \citep{Navarro1996} with a virial radius of ${R_\mathrm{vir} = 200\,\mathrm{kpc}}$ and concentration parameter $c = 12$, at a distance from the galactic centre of $R_\mathrm{D} = 8\,\mathrm{kpc}$ as in \citet{Li2017}.

With each simulation, we increase the level of stellar feedback complexity. The run labelled as \textit{S} only includes the feedback of SNe at the end of the lifetime of each massive star. In run \textit{SW}, we add continuous stellar wind feedback, in run \textit{SR}, we add ionising radiation from the massive stars, and in run \textit{SWR}, we account for the three feedback mechanisms together. The injection of CRs through SN remnants is introduced in the runs \textit{SWC} and \textit{SWRC}. All simulations cover 100 Myr of evolution. In this time the ISM can be evolved through multiple cycles of star formation. An overview of the simulations is given in Table \ref{tab:runs}. 

In Fig. \ref{fig:ov} we give a general overview how our most realistic simulation looks like. We show \textit{SWRC} at a later stage of its evolution at ${t = 65\,\mathrm{Myr}}$. The upper row is an edge-on view and the lower row a face-on view of the total gas column density, temperature as a slice, ionised, atomic, and molecular hydrogen column density, density-weighted magnetic field strength column, and CR energy density as a slice. The white circles in the \nth{1} and \nth{3} panels show the star clusters with their drawn size scaled to their respective masses. Please note that this size does not accurately reflect the actual physical size of stars clusters, which is only several parsecs \citep{McLaughlin2005, Bastian2013} and could not be properly visualised here. Translucent circles represent old star clusters with no active massive stars within them.

\begin{table}
    \caption{List of simulations with the included feedback processes. S: Type II SNe implemented as thermal energy input. W: stellar winds implemented as radial momentum and mass injection. R: ionising UV radiation (HII regions). C: injection and transport of non-thermal CRs (10 per cent of the SN energy) at SN explosion sites.}
    \begin{tabular}{lcccc}
        \hline
        Name & \textbf{S}upernovae & Stellar \textbf{W}inds & \textbf{R}adiation & \textbf{C}osmic rays  \\
        \hline
        \textit{S} & \checkmark & $\times$ & $\times$ & $\times$ \\
        \textit{SW} & \checkmark & \checkmark & $\times$ & $\times$ \\
        \textit{SWC} & \checkmark & \checkmark & $\times$ & \checkmark \\
        \textit{SR} & \checkmark & $\times$ & \checkmark & $\times$ \\
        \textit{SWR} & \checkmark & \checkmark & \checkmark & $\times$ \\
        \textit{SWRC} & \checkmark & \checkmark & \checkmark & \checkmark \\
        \hline
    \end{tabular}
    \label{tab:runs}
\end{table}

\section{Morphology and global evolution}\label{sec:morphology}

\begin{figure*}
    \centering
    \includegraphics[width=.8\linewidth]{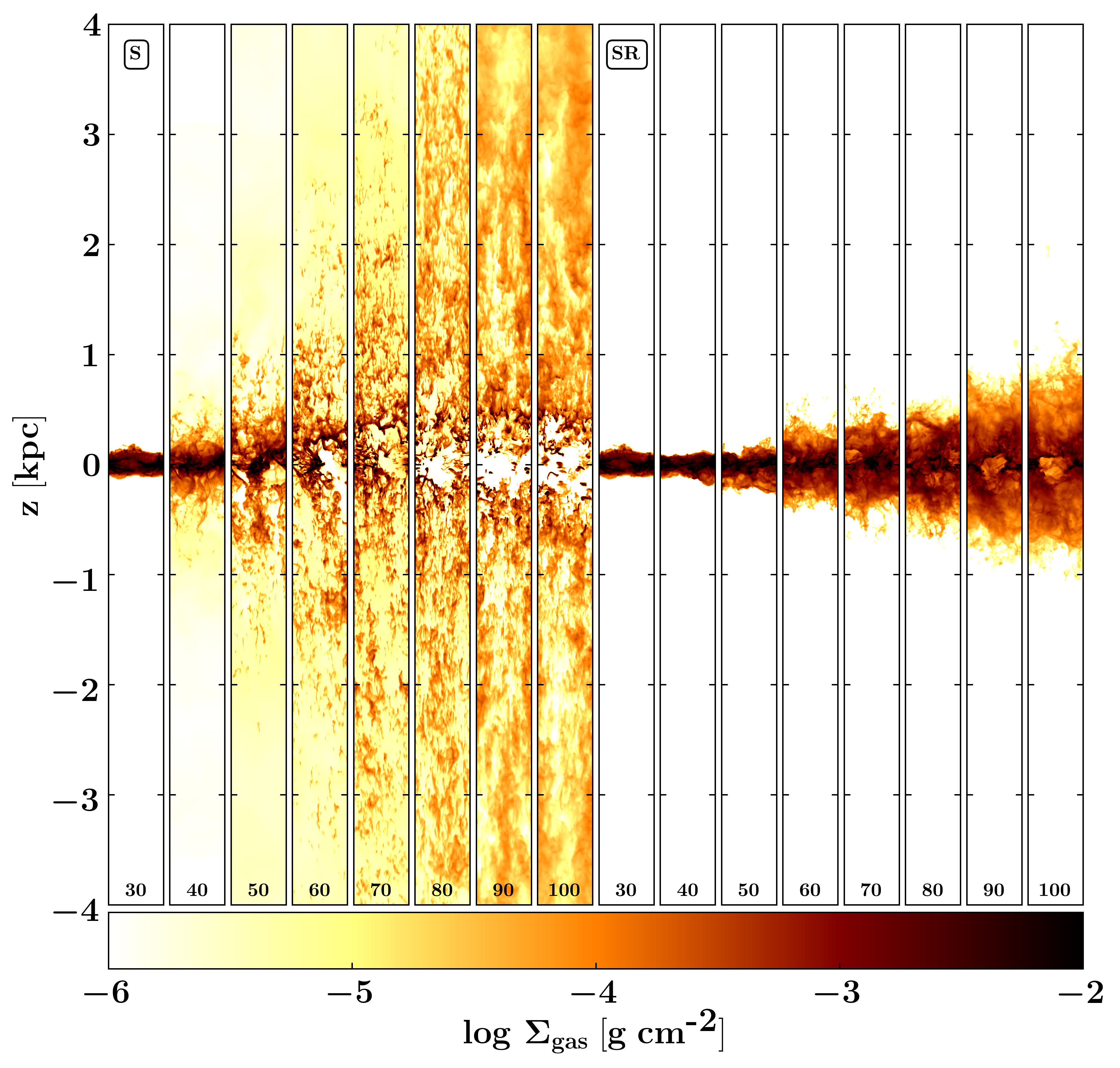}
    \caption{Edge-on view of the time evolution of the total gas column density for model \textit{S} with SN-only feedback and \textit{SR} with added ionising radiation from 30 to 100 Myr (indicated by the number at the bottom of each panel). The initial collapse triggers a starburst with strongly clustered SNe driving a highly structured outflow resulting in the dispersal of the mid-plane ISM. In longer-term evolution simulations, such a configuration might settle into a new equilibrium \citep{Kim2017}. The initial starburst is not inevitable. In our setup, it disappears if further early feedback processes from massive stars like stellar winds and radiation are included (see Fig. \ref{fig:evol_1} and Fig. \ref{fig:evol_2}).}
    \label{fig:evol_0}
\end{figure*}

\begin{figure*}
    \centering
    \includegraphics[width=.8\linewidth]{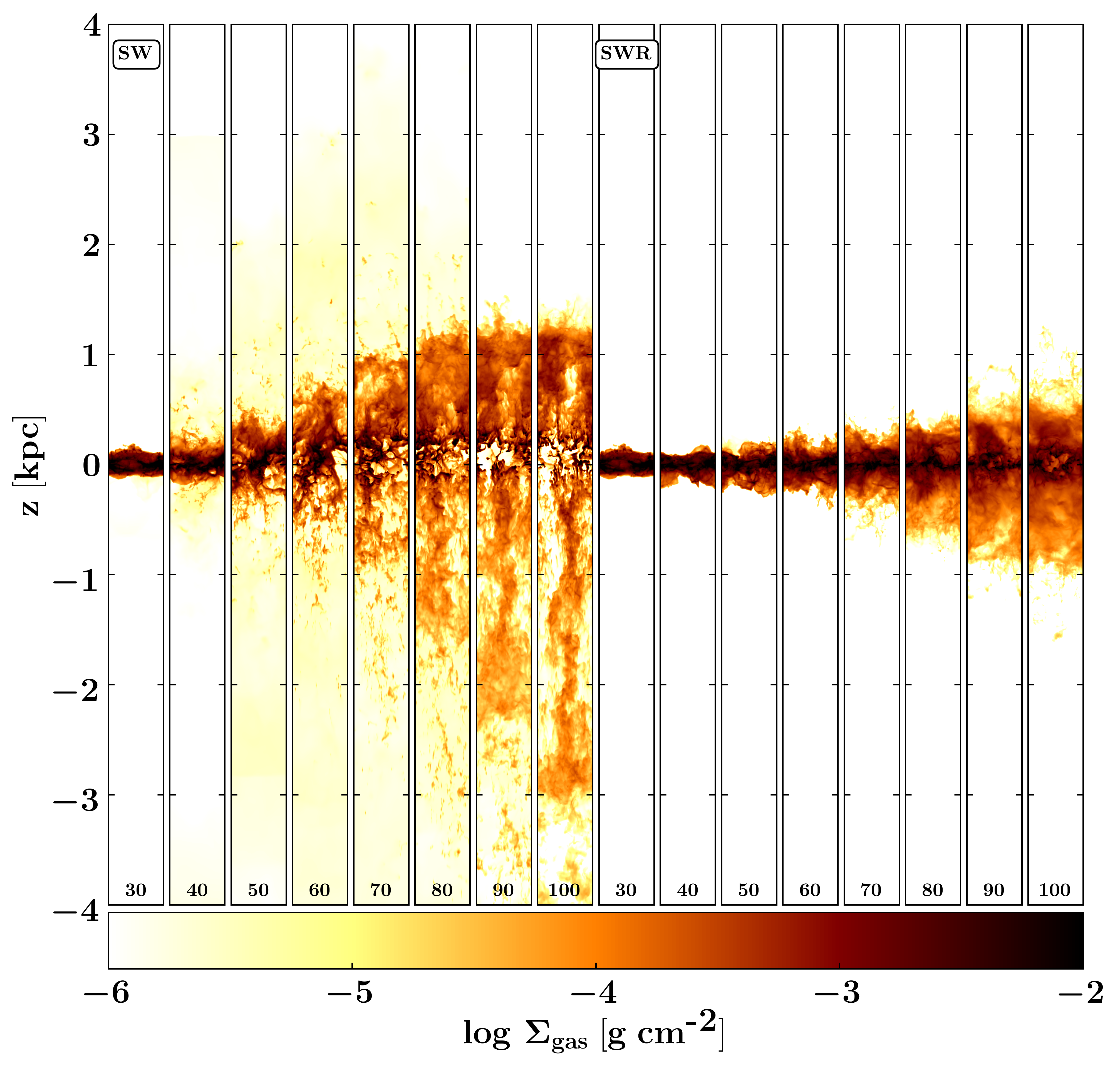}
    \caption{Column density time evolution for the models with added stellar winds (left, \textit{SW}) and added winds and ionising radiation (right, \textit{SWR}) from 30 to 100 Myr (see \ref{fig:evol_0} for the SN-only and SN-radiation models). The inclusion of stellar winds (\textit{SW}) reduces star formation (see Fig. \ref{fig:sfr}) by limiting accretion onto sink particles. Including ionising radiation (\textit{SWR}) regulates star formation even more and only weak outflows are driven.}
    \label{fig:evol_1}
\end{figure*}

\begin{figure*}
    \centering
    \includegraphics[width=.8\linewidth]{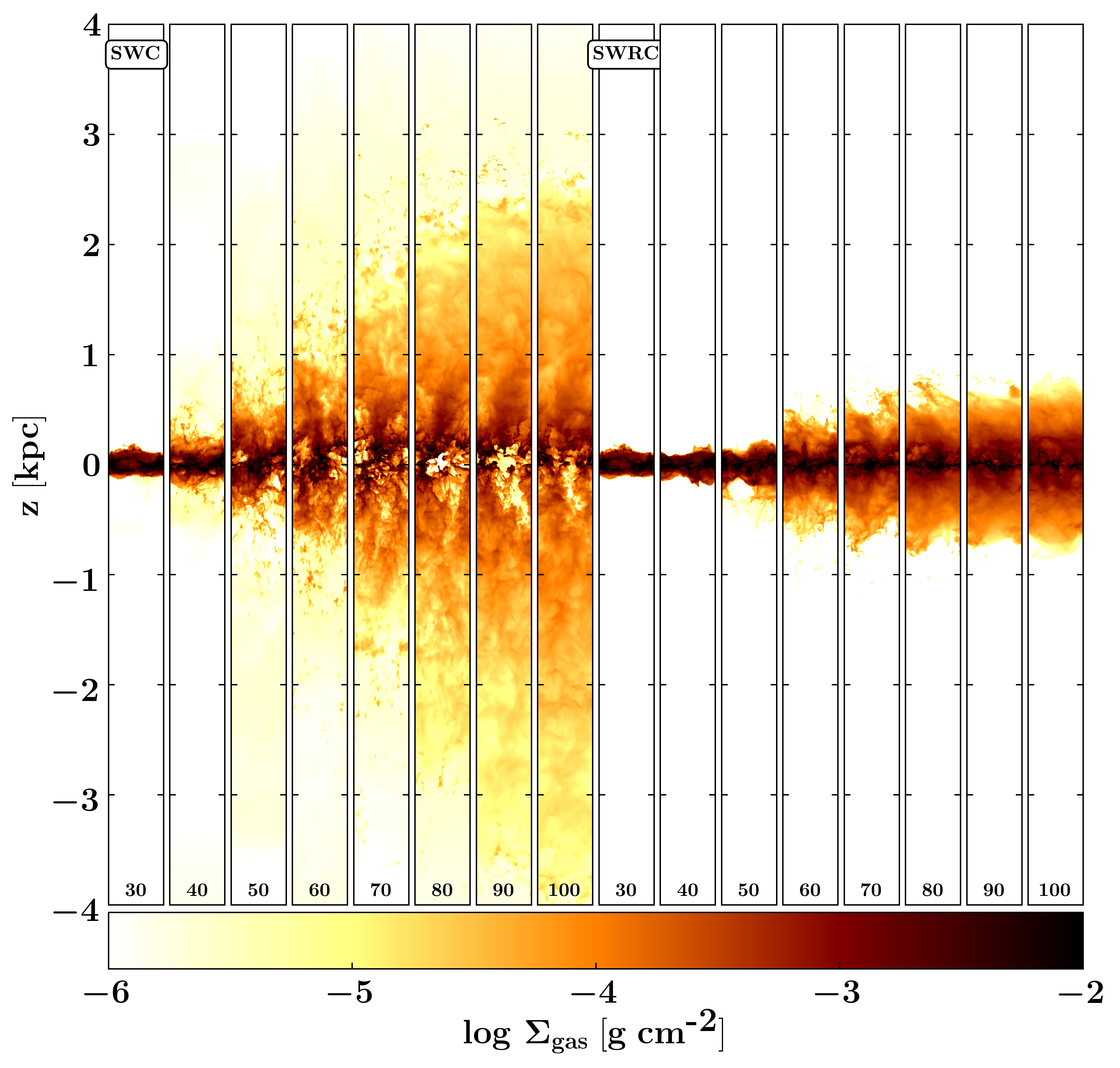}
    \caption{Column density time evolution for the models with added winds and CRs (left, \textit{SWC}) and added winds, ionising radiation, and CRs (right, \textit{SWRC}) from 30 to 100 Myr. The additional CR pressure smooths out the outflow. Overall, CRs have little effect on the mid-plane ISM in the early stages of the simulation but smooth out the gas in the outflow.}
    \label{fig:evol_2}
\end{figure*}

In Fig. \ref{fig:evol_0}, \ref{fig:evol_1} and \ref{fig:evol_2}, we show the time evolution of the total gas surface density $\Sigma_\mathrm{gas}$ seen edge-on for the three models without radiation \textit{S}, \textit{SW}, and \textit{SWC} on the left and the corresponding counterparts including radiation on the right. Models \textit{S} and \textit{SR} are depicted in Fig. \ref{fig:evol_0}, Fig. \ref{fig:evol_1} shows models \textit{SW} and \textit{SWR}, and Fig. \ref{fig:evol_2} models \textit{SWC} and \textit{SWRC}. We present the evolution from $t = 30\,\mathrm{Myr}$ until the end of the simulations at $t = 100$ Myr for for the full computational domain (${0.5\,\mathrm{kpc}\times 0.5\,\mathrm{kpc} \times \pm 4\,\mathrm{kpc}}$). Star formation starts after $\sim25\,\mathrm{Myr}$.

The models without radiation drive the strongest outflows (left panels in Fig. \ref{fig:evol_0}, \ref{fig:evol_1} and \ref{fig:evol_2}) due to an up to one order of magnitude higher SFR compared to their radiation counterparts (see Sec. \ref{sec:sfr}). Those strong outflows can even lead to a nearly complete depletion of gas in the mid-plane like in model \textit{S} (Fig. \ref{fig:evol_0} left). Ionising UV radiation prevents the star clusters from accreting more gas as soon as the first stars are born, resulting in a strong regulation of star formation \citep[see also][for similar conclusions]{Peters2017, Butler2017, Haid2018}. This reduced SFR results in weaker outflows launched at later stages (see Sec. \ref{sec:outflow}). CRs have a visible impact on the outflow structure during the first 100 Myr (\textit{SWC}, left of Fig. \ref{fig:evol_2} and \textit{SWRC}, right of Fig. \ref{fig:evol_2}) resulting in a smoother gas distribution \citep[see also][]{Simpson2016, Girichidis2016, Girichidis2018}. On this short time-scale, the additional CR pressure gradient does not result in significantly enhanced outflows as it requires some time to build up. On longer time-scales, CRs can become the dominant outflow driving mechanism as shown in \citet{Girichidis2016}. We will present the long-term evolution of simulations \textit{SWR} and \textit{SWRC} in a follow-up study (Rathjen et al., in prep.).

\section{Towards a complete model of the ISM}\label{sec:feeback}
\subsection{Star formation}\label{sec:sfr}

\begin{figure*}
	\centering
	\includegraphics[width=.99\linewidth]{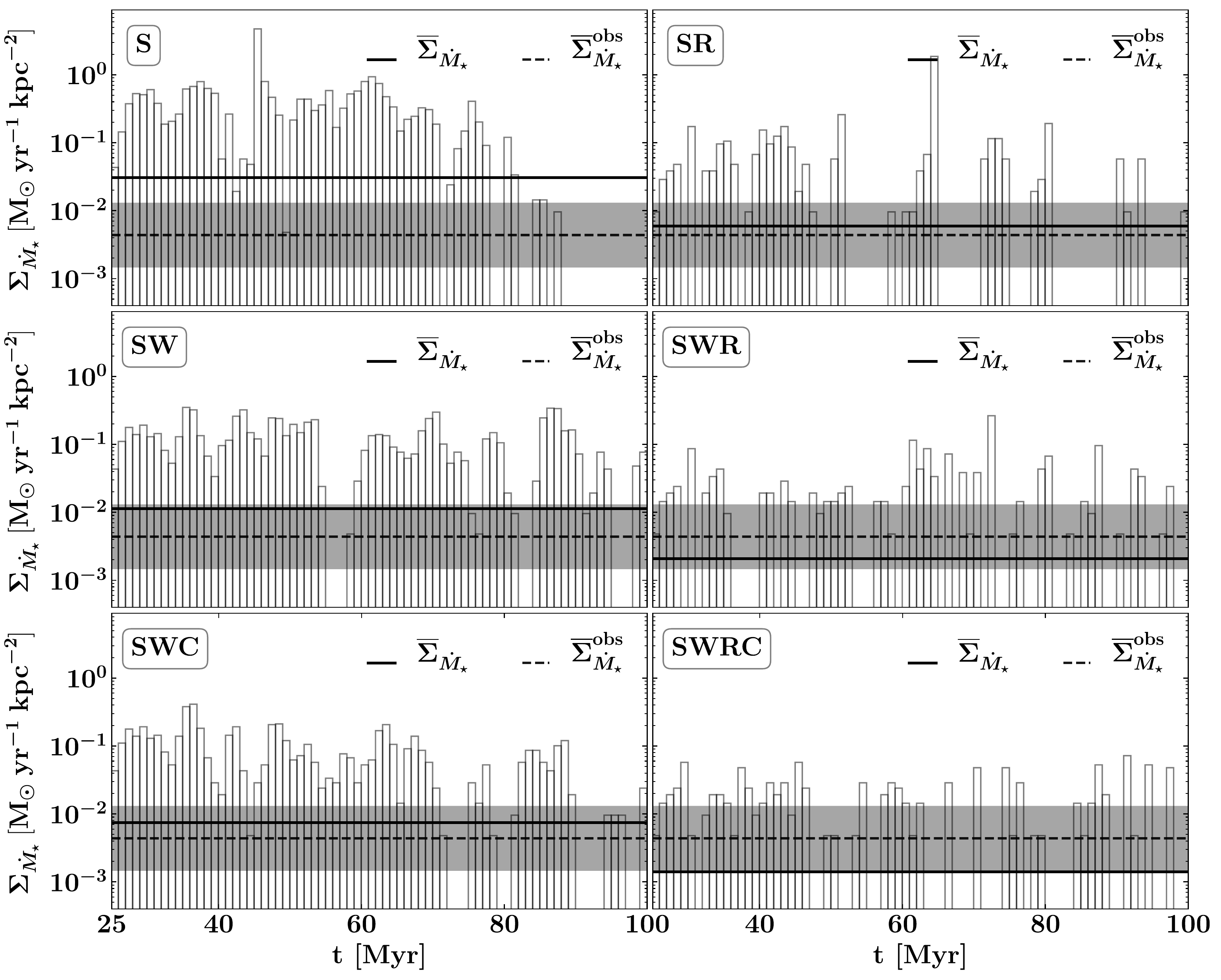}
	\caption{SFR surface densities $\Sigma_{\dot{M}_\star}$ for the different runs. The grey histograms indicate the instantaneous values. The solid line $\overline{\Sigma}_{\dot{M}_{\star}}$ is the time-averaged SFR surface density from $t = 25-100\,\mathrm{Myr}$ and the dashed black line is a mean SFR surface density for $\Sigma_\mathrm{gas} = 5-10\,\mathrm{M}_\odot\,\mathrm{pc}^{-2}$ from \citet{Leroy2008} (see Fig. \ref{fig:kennicutt}), including a factor of 3 uncertainty shaded in light grey. The models without radiation undergo an initial starburst, whereas \textit{SR}, \textit{SWR} and \textit{SWRC} (right panels) have lower SFRs comparable to observational estimates.\label{fig:sfr}}
\end{figure*}

In Fig. \ref{fig:sfr}, we show the SFR surface densities $\Sigma_{\dot{M}_\star}$ over time for the six models. The grey histograms indicate the instantaneous SFR surface densities, i.e. gas mass transformed into new stars, $\dot{M}_\mathrm{sink}$, in cluster $i$ in a period of $\Delta t = 1$ Myr per kpc$^2$:
\begin{equation}
    \Sigma_{\dot{M}_{\star}}(t) = \frac{1}{A} \sum^{N_\mathrm{sink}}_{i=1} \dot{M}_\mathrm{sink, i},
\end{equation} for $t-\frac{\Delta t}{2} < t < t + \frac{\Delta t}{2}$ and the surface area of the mid-plane ISM $A = 0.25\,\mathrm{kpc}^2$.

\begin{figure}
	\centering
	\includegraphics[width=.99\linewidth]{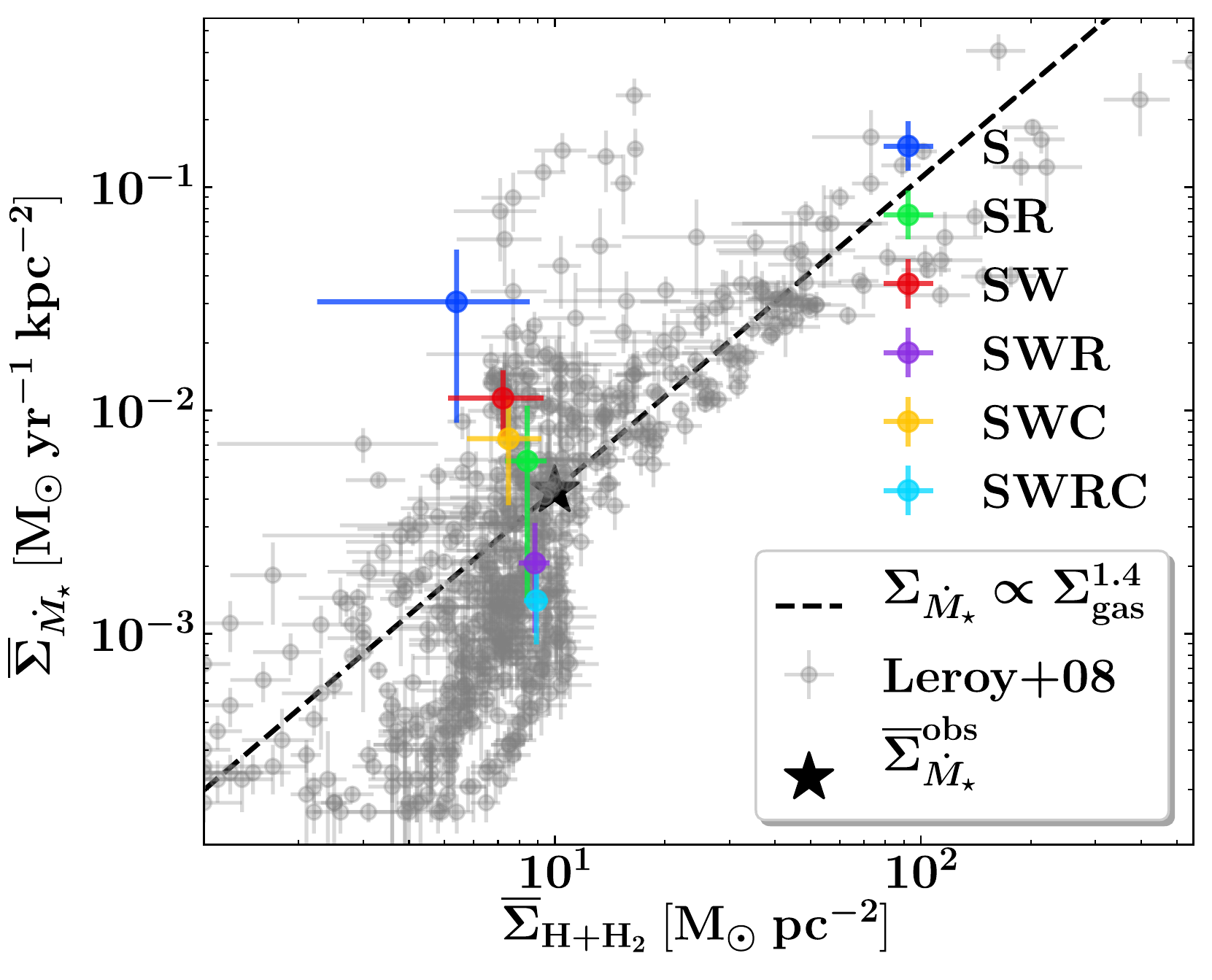}
	\caption{SFR surface densities vs. gas surface densities of the atomic and molecular hydrogen gas in the disc ($z = \pm 250\,\mathrm{pc}$), averaged over $t = 25-100\,\mathrm{Myr}$ with 1$\sigma$ scatter. The black dashed line indicates a Kennicutt-Schmidt relation \citep{KennicuttJr.1998} slope of $\Sigma_{\dot{M}_\star} \propto \Sigma_{\mathrm{gas}}^{1.4}$. The grey dots are observational data from spatially resolved patches of nearby star-forming galaxies \citep{Leroy2008}. The black star indicates an average star formation rate value for gas surface densities $\Sigma_\mathrm{gas} = 5-10\,\mathrm{M}_\odot\,\mathrm{pc}^{-2}$ of $\overline{\Sigma}^\mathrm{obs}_{\dot{M}_\star} = 4.4 \times 10^{-3}\,\mathrm{M}_\odot\,\mathrm{yr}^{-1}\,\mathrm{kpc}^{-2}$. The SN-only run (\textit{S}) has a very high SFR for its gas surface density. Models including winds (\textit{SW}, \textit{SWC}) and, in particular, radiation (\textit{SR}, \textit{SWR}, \textit{SWRC}) are closer to most observational values.}
	\label{fig:kennicutt}
\end{figure}

The dashed line in each panel indicates a fiducial SFR surface density using the mean value of the \citet{Leroy2008} data for a H$_2$ + H gas surface density range of $\Sigma_\mathrm{gas} = 5-10\,\mathrm{M}_\odot\,\mathrm{pc}^{-2}$ (see Fig. \ref{fig:kennicutt} for more details). We indicate a factor 3 scatter by the shaded area. The solid black line $\overline{\Sigma}_{\dot{M}_{\star}}$ is the mean value of the SFR surface density averaged from $t = 25-100\,\mathrm{Myr}$ for each simulation respectively.

For the different models the average SFR surface densities $\overline{\Sigma}_{\dot{M}_\star}$ with 1$\sigma$ scatter are:

\begin{tabular}{lll}
        \textit{S}: & $\overline{\Sigma}_{\dot{M}_\star} = (3.06 \pm 2.18) \times 10^{-2}$ & $\mathrm{M}_\odot\,\mathrm{yr}^{-1}\,\mathrm{kpc}^{-2}$ \\
        \textit{SW}: & $\overline{\Sigma}_{\dot{M}_\star} = (1.13 \pm 0.38) \times 10^{-2}$ & $\mathrm{M}_\odot\,\mathrm{yr}^{-1}\,\mathrm{kpc}^{-2}$ \\
        \textit{SWC}: & $\overline{\Sigma}_{\dot{M}_\star} = (7.46 \pm 3.70) \times 10^{-3}$ & $\mathrm{M}_\odot\,\mathrm{yr}^{-1}\,\mathrm{kpc}^{-2}$ \\
        \textit{SR}: & $\overline{\Sigma}_{\dot{M}_\star} = (5.93 \pm 4.54) \times 10^{-3}$ & $\mathrm{M}_\odot\,\mathrm{yr}^{-1}\,\mathrm{kpc}^{-2}$ \\
        \textit{SWR}: & $\overline{\Sigma}_{\dot{M}_\star} = (2.07 \pm 1.06) \times 10^{-3}$ & $\mathrm{M}_\odot\,\mathrm{yr}^{-1}\,\mathrm{kpc}^{-2}$ \\
        \textit{SWRC}: & $\overline{\Sigma}_{\dot{M}_\star} = (1.41 \pm 0.51) \times 10^{-3}$ & $\mathrm{M}_\odot\,\mathrm{yr}^{-1}\,\mathrm{kpc}^{-2}$.\\
\end{tabular}

The model with SN-only, \textit{S}, rapidly evolves into a starburst with an average SFR about one order of magnitude above the observationally motivated value of $\overline{\Sigma}^\mathrm{obs}_{\dot{M}_\star} = 4.4 \times 10^{-3}\,\mathrm{M}_\odot\,\mathrm{yr}^{-1}\,\mathrm{kpc}^{-2}$ (dashed black lines in Fig. \ref{fig:sfr}) for a gas surface density range of $\Sigma_\mathrm{gas} = 5-10\,\mathrm{M}_\odot\,\mathrm{pc}^{-2}$ \citep{Leroy2008}. We chose this range in gas surface density because it represents the upper and lower limits of average gas surface densities in our simulations. Due to the lack of early feedback processes from massive stars, gas can be accreted by the cluster sinks until the first SNe explode \citep{Gatto2017, Peters2017}. For this simulation, the median cluster mass is $M_\mathrm{median} = 1.6 \times 10^{4}$ M$_\odot$ with an average number of $\overline{N}_\star = 184$ massive stars per cluster (see Table \ref{tab:cluster}). The highly clustered SNe drive a strong outflow and the mid-plane star-forming ISM completely disperses (see Fig. \ref{fig:evol_0}, left panel). Therefore star formation is terminated by the depletion and dispersal of the cold gas reservoir during the last $\sim15\,\mathrm{Myr}$ of model \textit{S}.  The SFR of model \textit{SW} is lower than in model \textit{S} by about a factor of 3. All three non-radiation models lie above the observationally motivated value.

For the radiation runs \textit{SR}, \textit{SWR} and \textit{SWRC} the behaviour is qualitatively different. The SFR surface density is about a factor of 5 lower than for the respective runs without radiation and agrees with observational expectations. While the initial starburst is already slightly suppressed in models \textit{SW} and \textit{SWC} by the early feedback in form of stellar winds, it is absent in the radiation runs \textit{SR}, \textit{SWR} and \textit{SWRC}. Comparing \textit{SW}, \textit{SR} and \textit{SWR}, the SFR drops by nearly one order of magnitude when adding the radiation but only by about a factor of $\sim2.5$ when adding winds. Thus, ionising UV radiation seems more important for quenching the SFR than stellar winds, at least for the models at $\sim4\,\mathrm{pc}$ resolution presented here. Those findings qualitatively agree with earlier studies \citep{Butler2017, Peters2017} on this topic and also higher resolution simulations on smaller scales \citep[see e.g.][]{Dale2014, Geen2015, Geen2017, Haid2018}. CRs do not directly impact the gas structure of the disc and the accretion behaviour of the sink particles. The ISM is still dominated by the thermal and kinetic gas pressures and the strongest impact of the CRs is seen only in the outflow region.

In Fig. \ref{fig:kennicutt}, we compare the gas surface densities $\Sigma_\mathrm{H+H_2}$ and SFR surface densities $\Sigma_{\dot{M}_\star}$ of our models with spatially resolved observations from local star-forming spiral- and dwarf-galaxy patches (light grey dots, \citet{Leroy2008}). The dashed black line is the Kennicutt-Schmidt relation (KS-relation) $\Sigma_{\dot{M}_\star} \propto \Sigma^{1.4}_\mathrm{H+H_2}$ \citep{KennicuttJr.1998} centred on the average SFR surface density for gas surface densities $\Sigma_\mathrm{gas} = 5-10\,\mathrm{M}_\odot\,\mathrm{pc}^{-2}$ with $\overline{\Sigma}^\mathrm{obs}_{\dot{M}_\star} = 4.4 \times 10^{-3}\,\mathrm{M}_\odot\,\mathrm{yr}^{-1}\,\mathrm{kpc}^{-2}$. This range in gas surface densities is chosen because it represents the upper and lower limits of the averaged gas surface densities in our simulations. Shown are the averaged values of our models from $t = 25-100\,\mathrm{Myr}$, with the error bars indicating 1$\sigma$ scatter. Including early feedback processes in the form of stellar winds (\textit{SW}) and, in particular, radiation (\textit{SR}) reduces the SFRs, resulting in values more consistent with the mean value derived from observations \citep{Leroy2008}. CRs (\textit{SWC} and \textit{SWRC}) have a weak additional impact and only slightly reduce the SFR. We note that around a gas surface density of $\Sigma_\mathrm{gas} \sim10 \,\mathrm{M}_\odot\,\mathrm{pc}^{-2}$ the observations show an enormous range of star formation rates covering $\sim3$ orders of magnitudes. Therefore models for higher surface densities might provide stronger physical constraints \citep[see e.g.][]{Gong2020}. 

\begin{figure}
	\centering
	\includegraphics[width=.99\linewidth]{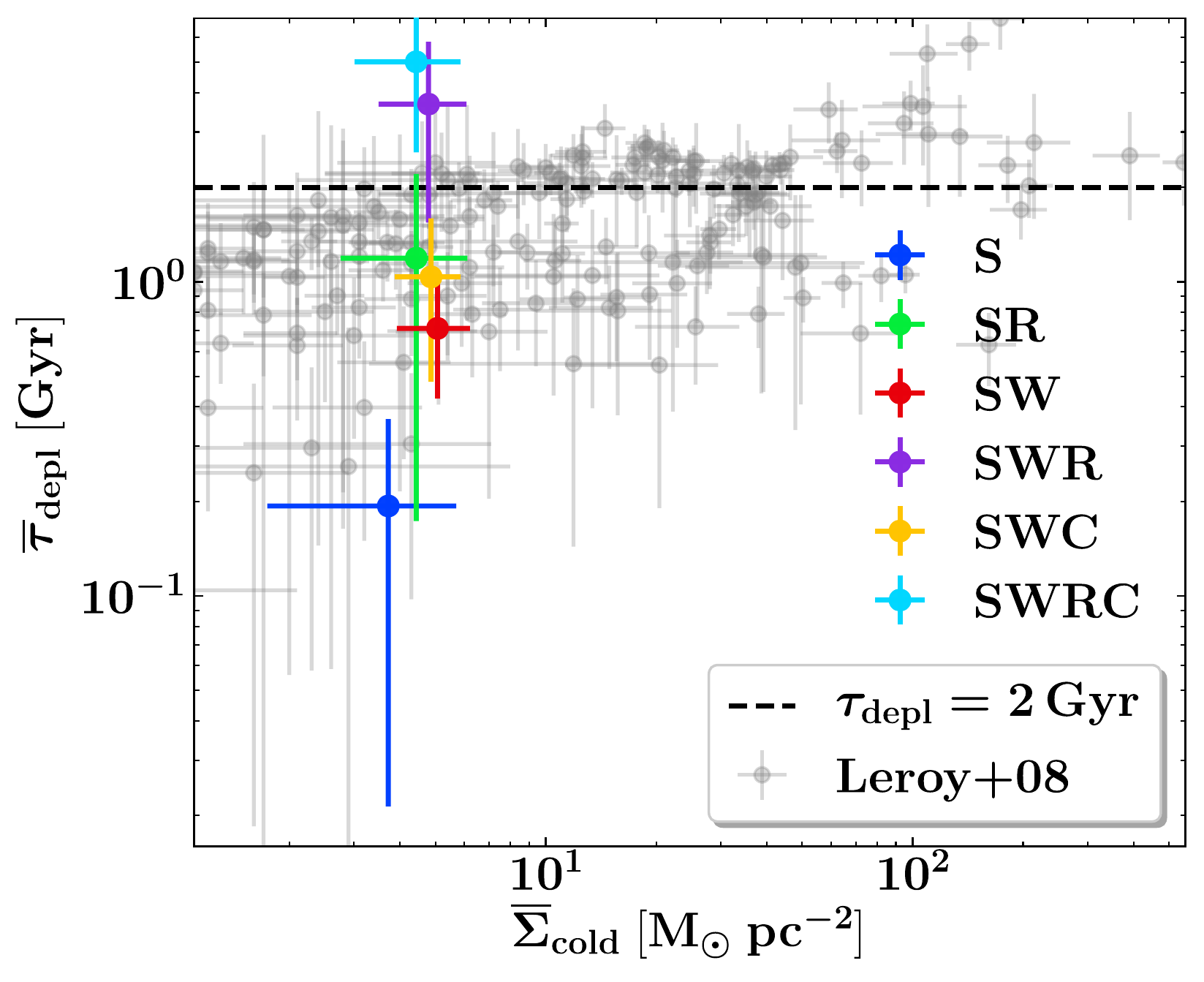}
	\caption{Average gas depletion times $\overline{\tau}_\mathrm{depl}$ of the cold gas phase ($T < 300 \,\mathrm{K}$) vs. average gas surface densities of the cold phase $\overline{\Sigma}_\mathrm{cold}$ with 1$\sigma$ scatter. The time averages are taken from $t = 25-100 \,\mathrm{Myr}$. The horizontal dashed line indicates a constant depletion time of $\tau_\mathrm{depl} = 2\,\mathrm{Gyr}$. The grey dots are observed depletion times for molecular gas H$_2$ from \citet{Leroy2008}.}
	\label{fig:bigiel}
\end{figure}

In Fig. \ref{fig:bigiel} we show the average depletion times $\overline{\tau}_\mathrm{depl} = \overline{\Sigma}_\mathrm{cold} \times \overline{\Sigma}_{\dot{M}_\star}^{-1}$ of the simulated cold gas phase ($T < 300 \,\mathrm{K}$) against the average cold phase gas surface density $\overline{\Sigma}_\mathrm{cold}$ in the mid-plane $z = \pm 250\,\mathrm{pc}$. The observational data from \citet{Leroy2008} shows gas depletion times for molecular H$_2$ gas. The dashed black line indicates a constant depletion time of $\tau_\mathrm{depl} = 2\,\mathrm{Gyr}$ as favoured by observations \citep{Bigiel2008}.

For our models, we find depletion times ranging from $0.19\,\mathrm{Gyr}$ in \textit{S} to $5.02\,\mathrm{Gyr}$ in \textit{SWRC}. The two other models including radiation \textit{SR} and \textit{SWR} exhibit average depletion times of $1.19\,\mathrm{Gyr}$ and $3.68\,\mathrm{Gyr}$, respectively. The two wind models without radiation (\textit{SW} and \textit{SWC}) have cold gas depletion times around $0.7-1.0\,\mathrm{Gyr}$. A constant molecular gas depletion time, as suggested by the observations, informs of a linear relationship between the molecular gas surface density $\Sigma_\mathrm{H_2}$ and SFR surface density $\Sigma_{\dot{M}_\star}$, i.e. a constant efficiency of transforming molecular gas into stars. \citet{Bigiel2008} find a constant molecular gas depletion time of $\tau_\mathrm{depl} = 2\,\mathrm{Gyr}$ with a rms scatter of $0.8\,\mathrm{Gyr}$ for a sample of 18 nearby galaxies, measured over a $\Sigma_\mathrm{H_2}$ range of $\sim3-50\,\mathrm{M}_\odot\,\mathrm{pc}^{-2}$ (shown as black dashed line in Fig. \ref{fig:bigiel}). Our most realistic models including early feedback from ionising UV radiation and stellar winds (\textit{SWR} and \textit{SWRC}) lie remarkably close to the constant depletion time of 2 Gyr inferred by observations. Our SN-only model \textit{S} shows a depletion time of $\sim200\,\mathrm{Myr}$ at the lower bound of the observational scatter. In \textit{S}, star formation is fully quenched after $\sim90\,\mathrm{Myr}$ (see Fig. \ref{fig:sfr}) and the gas reservoir is completely used up at later stages (see also the \textit{holes} in the mid-plane gas column density in the left panel of Fig. \ref{fig:evol_0}). This indicates that the SFR is regulated by the galactic outflow, instead of depletion of the cold gas phase via star formation.

We show the depletion times for the cold gas instead of presenting it for the H$_2$, which is included in our chemical network. The reason for this is that the H$_2$ formation is likely not fully converged at our spatial resolution of $\Delta x \approx 4 \,\mathrm{pc}$. The cold gas phase is the regime where molecular gas would form and is used as a proxy for estimating the molecular gas surface density $\Sigma_\mathrm{H_2}$. We might be over-estimating $\Sigma_\mathrm{H_2}$ with this assumption, which would hence result in a too large estimate for the depletion times

\subsection{Star cluster properties}\label{sec:cluster}

\begin{figure}
	\centering
	\includegraphics[width=.99\linewidth]{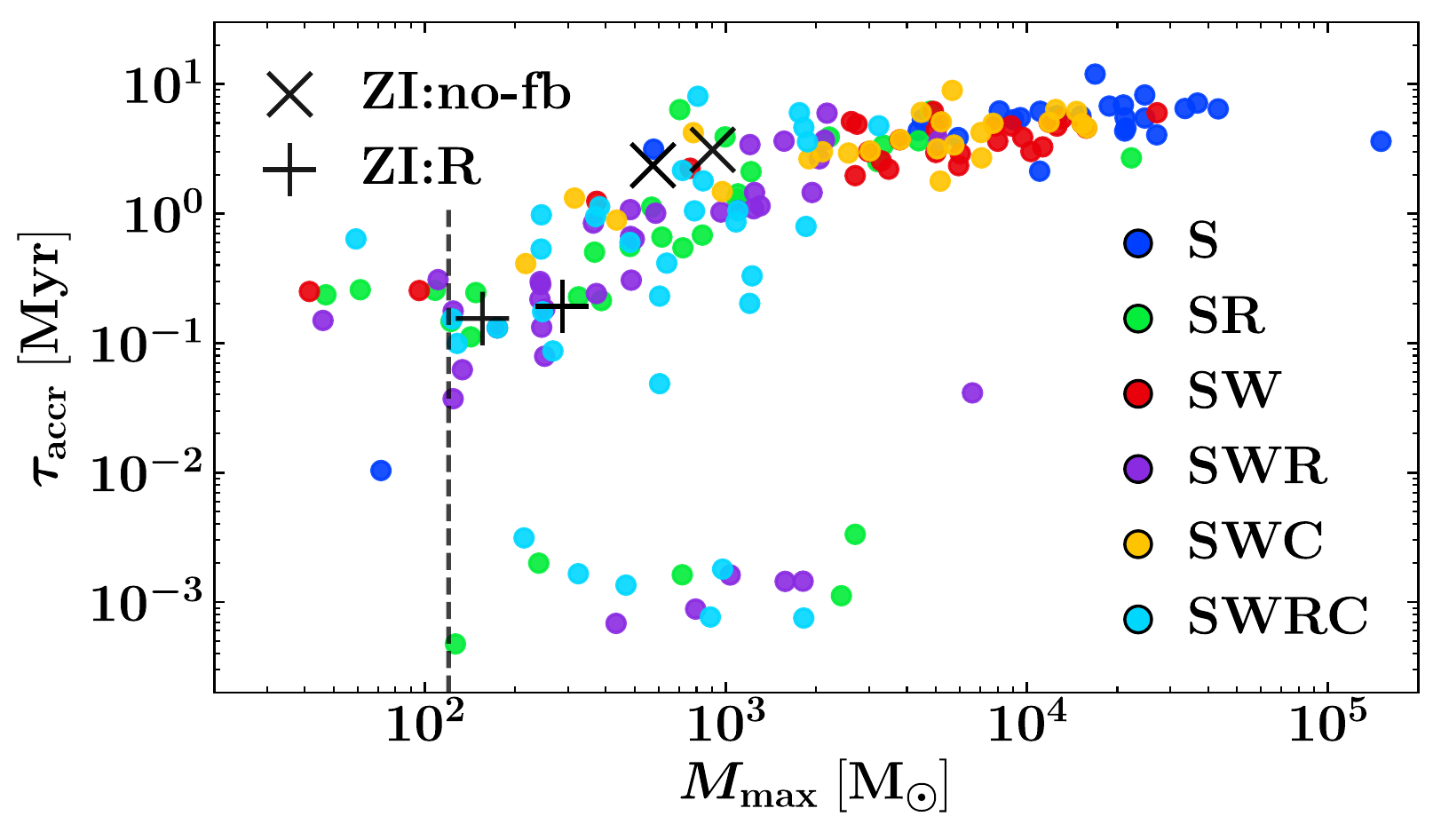}
	\caption{Maximum accretion time-scale $\tau_\mathrm{accr}$ of the star cluster sink particles against the maximum accreted mass $M_\mathrm{max}$ of each cluster sink in the simulations. The black crosses are from molecular cloud zoom-in simulations with a resolution of $\Delta x = 0.112\,\mathrm{pc}$ \citep{Haid2019}, incorporating the same methods for radiative transfer, chemistry and sink particle creation as we do in this study.  Without ionising radiation feedback, a slightly lower number of significantly more massive clusters are formed, with longer accretion times (see also Table \ref{tab:cluster}). The dashed black vertical line indicates a cluster mass of $120\,\mathrm{M}_\odot$. We form a handful of star cluster sink particles with $M_\mathrm{max} < 120\,\mathrm{M}_\odot$. Those clusters do not host massive stars and do not contribute feedback. Inherently, there is no lower mass limit for the star cluster sink particles we form.}
	\label{fig:accr}
\end{figure}

Massive stars in galaxies are believed to form hierarchically, embedded in dense molecular clouds and young massive clusters \citep{Lada2003, PortegiesZwart2010, Grasha2017}.
Fig. \ref{fig:accr} shows the accretion properties of the star cluster sink particles formed in our simulations. We show the maximum accretion time-scale $\tau_\mathrm{accr}$ defined as the time each cluster takes to reach its respective maximum mass $M_\mathrm{max}$ through accretion against $M_\mathrm{max}$. Lower mass clusters ($M_\mathrm{max} < 120\,\mathrm{M_\odot}$, indicated by the dashed vertical line in Fig. \ref{fig:accr}) do not accrete enough gas to form massive stars and have no active feedback channel. Their accretion properties are solely determined by the availability of gas in their natal environment. 
The data-points indicated by crosses come from a high-resolution simulation done by \citet{Haid2019}, which are part of the \textsc{SILCC-Zoom} project \citep[see e.g.][for details about the zoom-in simulations]{Seifried2017}. \citet{Haid2019} take two self-consistently formed molecular clouds (MC) from the first set of the \textsc{SILCC} simulation suite \citep{Walch2015, Girichidis2016} which have been identified in \citet{Seifried2017} and re-calculate the central part (a cube with side length $l = 40\,\mathrm{pc}$) at a resolution of $\Delta x \approx 0.122 \,\mathrm{pc}$. They run two sets of simulations for each identified MC one without any form of feedback (labelled here as ZI:no-fb), equivalent to our model \textit{S}, and one with ionising UV radiation (labelled here as ZI:R), equivalent to our model \textit{SR}.
We group the stars and sub-clusters formed in each MC together as one cluster and plot the mass-weighted average accretion time against the average maximum mass of those clusters. Being part of the same framework, \citet{Haid2019} use the same methods for radiative transfer (\textsc{TreeRay} W\"unsch et al., submitted), the same time-dependent chemical network, including heating and cooling, and the same sink particle creation and accretion mechanisms, albeit with different parameters. Their accretion radius is $r_\mathrm{accr} = 0.31\,\mathrm{pc}$ (corresponding to $2.5 \times \Delta x$) and their density threshold is $n_\mathrm{sink} \approx 5 \times 10^3\,\mathrm{cm}^{-3}$.

Stellar feedback has a strong influence on the overall formation of the star clusters.
Without continuous feedback (model \textit{S}), the clusters become significantly more massive but the total number of formed star clusters $N_\mathrm{cluster}$ is slightly lower than in models with wind and radiation.
Wind feedback can push down the median cluster mass $M_\mathrm{median}$ and the average number of massive stars in each cluster $\overline{N}_\star$ by a factor of $\sim3$. The strongest effect, however, is seen in the models including ionising radiation. The clusters are remarkably lighter with the most massive one in model \textit{SWRC} nearly 50 times less massive than in model \textit{S}. The average number of massive stars per cluster drops from $\overline{N}_\star = 184$ in simulation \textit{S} to only $\overline{N}_\star = 6$ in model \textit{SWRC}.
The maximum accretion time-scale $\tau_\mathrm{accr}$ is limited by the lifetime of the most massive star in a cluster after it undergoes a supernova explosion if no other feedback channels are included. Stellar winds only have a limited impact on the accretion time-scale. When radiative feedback is not included, the accretion time-scales do not depend on the total accreted mass, with only a few outliers. This trend is similar to results from higher resolution zoom simulations of individual molecular clouds \citep{Haid2019}.
Cosmic rays seem to not play a role in cluster formation since the differences between \textit{SW} and \textit{SWC}, as well as \textit{SWR} and \textit{SWRC} are negligible. Early stellar feedback strongly suppresses clustering \citep[see also recent results from][]{Hu2017, Smith2020}, which also inhibits the formation of the super-bubbles needed to generate a volume-filling hot gas phase as discussed in the next sections. Our result that radiative feedback plays the most crucial role in regulating cluster formation and star formation properties is also found in other studies by e.g. \citet{Murray2010, Dale2012, Howard2017, Peters2017}. We want to note that the effect of feedback does not change on smaller scales (compare with the \citet{Haid2019} data in Fig. \ref{fig:accr}) and the choice of sink particle accretion parameters does not qualitatively change the outcome. 
The properties of the star cluster sink particles, as well as the percentage of unresolved SNe with momentum injection $f_\mathrm{mom}$ are listed in Table \ref{tab:cluster}.

\begin{table}
    \caption{Star cluster sink properties of the six models. We list the mass of the most massive cluster sink formed $M_\mathrm{cl}$, the median mass of the formed cluster sinks when they stopped accreting $M_\mathrm{median}$, the median accretion time-scale of the cluster sinks $\tau_\mathrm{accr}$, the total number of formed cluster sinks $N_\mathrm{cluster}$, and the average number of massive stars per cluster sink $\overline{N}_\star$, computed as the total number of massive stars formed divided by the total number of cluster sinks formed. The percentage of unresolved SNe with momentum injection $f_\mathrm{mom}$ is given in the last column. Radiation feedback inhibits clustering and drastically reduces the median and maximum mass of the star cluster sinks, preventing the formation of super-bubbles and resulting in less effective SNe.}
    \begin{tabular}{lcccccc}
        \hline
        Run & $M_\mathrm{cl}$ & $M_\mathrm{median}$ & $\tau_\mathrm{accr}$  & $N_\mathrm{cluster}$ & $\overline{N}_\star$ & $f_\mathrm{mom}$\\
         & [$\mathrm{M}_\odot$] & [$\mathrm{M}_\odot$] & [Myr] & & & [\%]\\
        \hline
        \textit{S}    & $1.5 \times 10^{5}$ & $1.6 \times 10^{4}$ & 5.39 & 26 & 184 & 5.3\\
        \textit{SW}   & $2.7 \times 10^{4}$ & $5.1 \times 10^{3}$ & 3.62 & 30 &  59 & 6.9\\
        \textit{SWC}  & $1.6 \times 10^{4}$ & $5.1 \times 10^{3}$ & 3.79 & 24 &  48 & 7.5\\
        \textit{SR}   & $2.2 \times 10^{4}$ & $7.1 \times 10^{2}$ & 1.39 & 30 &  15 & 5.9\\
        \textit{SWR}  & $6.6 \times 10^{3}$ & $5.0 \times 10^{2}$ & 0.98 & 37 &   8 & 4.2\\
        \textit{SWRC} & $3.2 \times 10^{3}$ & $6.4 \times 10^{2}$ & 1.25 & 33 &   6 & 5.3\\
        \hline
    \end{tabular}
    \label{tab:cluster}
\end{table}

\begin{figure}
	\centering
	\includegraphics[width=.99\linewidth]{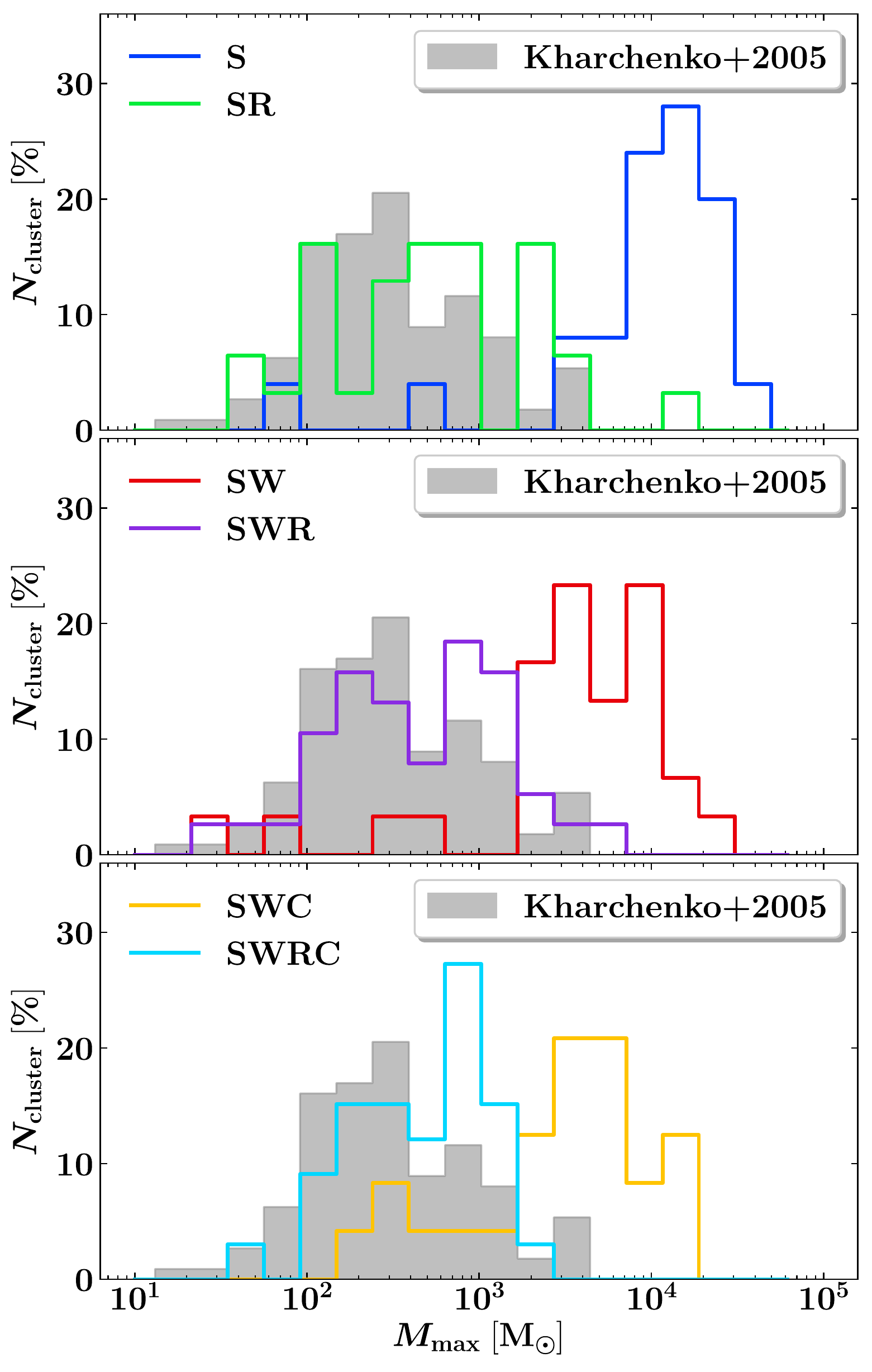}
	\caption{Histograms of the cluster sink mass distributions for the six models. For better readability, we split the depiction into three panels. The grey shades histograms come from observational data of 114 open star cluster in the solar neighbourhood ($d < 600\,\mathrm{pc}$) by \citet{Kharchenko2005}. Without the accretion limiting effects of ionising UV radiation, the star clusters grow to masses greater than $10^4\,\mathrm{M_\odot}$, atypical for the solar neighbourhood.}
	\label{fig:hist}
\end{figure}

In Fig. \ref{fig:hist} we show the star cluster sink mass distribution normalised to the total number of clusters formed in each model. To improve readability, we split the six models into three panels, each panel grouping together the respective models with and without ionising UV radiation. The grey shaded histograms represented observational data of 114 open clusters in the solar neighbourhood ($d < 600\,\mathrm{pc}$) taken out of a catalogue of 520 Galactic open clusters \citep{Kharchenko2005}. The sharp cut-off at the low mass end of the distribution suggests a complete sample for clusters more massive than $M \gtrsim 10^2\,\mathrm{M_\odot}$, however, this sample includes cluster with age estimates between a few Myr to a few $10^3$ Myr, way older than the total simulated time in our models. We do not incorporate any cluster disruption mechanisms in our simulations, so the comparison of our data to the observational data is mostly qualitative. The small number of clusters formed in our models ($N_\mathrm{cluster} \sim 30$) does not allow us to meaningfully sample a cluster mass function. Nonetheless, there is a clear trend of forming too massive clusters, atypical for the local solar neighbourhood, when omitting ionising UV radiation.

\subsection{The importance of supernova ambient densities}\label{sec:sn}

The ambient ISM densities at SN sites are of fundamental importance for their local and global dynamical and thermal impact \citep{Naab2017}. At high environmental densities, the imparted SN energy is rapidly cooled away and the energy and momentum coupling to the ambient gas is very low \citep{Gatto2015, Walch2015a, Kim2015, Haid2016}. For low ambient densities radiation losses are minor and super-bubbles \citep{MacLow1988, Wunsch2008} with a high hot gas VFF can be created by consecutive and spatially overlapping SN events \citep{MacLow1988, Creasey2013, Fielding2017}. This will significantly support the driving of outflows from the ISM \citep[see e.g.][]{Li2020}. Numerical experiments by \citet{Walch2015a} and \citet{Girichidis2016a} have shown that it makes a qualitative difference whether SNe at a fixed rate, i.e. with the same total energy input, explode at density peaks or random positions in the medium. Of course, ambient densities can be affected by the highly non-linear interaction of SNe, stellar winds, ionising radiation and clustering \citep{Kim2011, Hennebelle2014, Li2015, Walch2015a, Girichidis2016a, Gatto2017, Naab2017, Hu2017, Rahner2017, Fielding2018, Haid2018, Haid2019, Rahner2019, Smith2020}.

\begin{figure*}
\centering
\includegraphics[width=.99\linewidth]{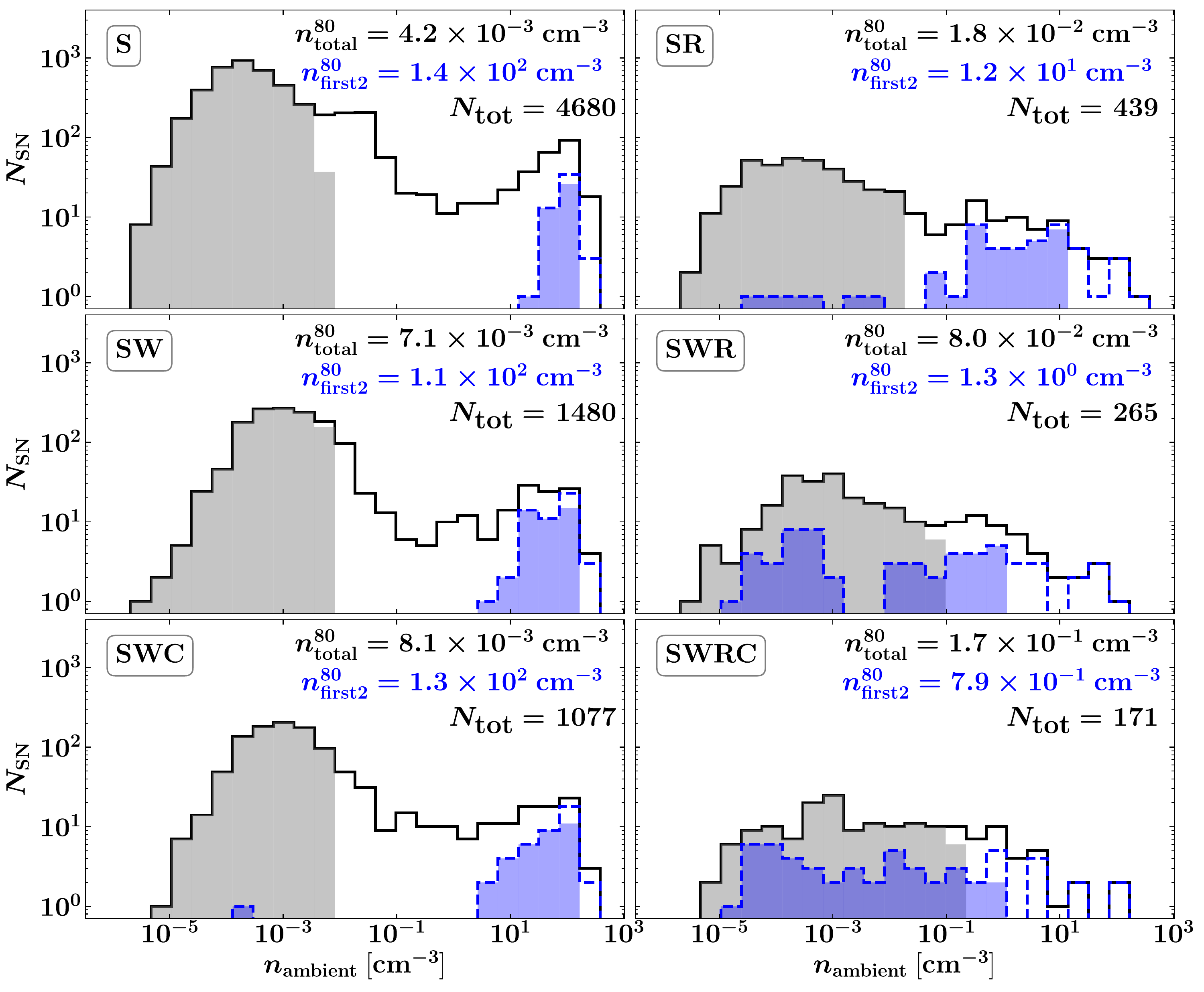}
	\caption{Distribution of the ambient gas densities at SN explosion sites. Shaded in grey are the histograms including 80 per cent of all SNe exploding at the lowest densities up to corresponding densities $n_\mathrm{total}^{80}$ given in the respective panels. The blue dashed histograms indicate the densities of the first two SNe in each star cluster sink, with the respective histograms including 80 per cent of the first two SNe per star cluster sink shaded in blue. For SN-only (\textit{S}) the ambient density distribution is bi-modal, which is a clear sign of strong clustering. Most SNe explode in low-density bubbles created by previous events, 80 per cent of all SNe explode at ambient densities below $\sim4 \times 10^{-3}\,\mathrm{cm}^{-3}$ (grey histogram). For runs without radiation (\textit{S}, \textit{SW}, \textit{SWC}) the first SNe of a cluster always explode at high ambient densities $\sim20-200\,\mathrm{cm}^{-3}$ (blue dashed histograms). The high-density peak disappears with the inclusion of radiation (\textit{SR}, \textit{SWR}, \textit{SWRC}, right panels) and the distributions become flatter. While most SNe now explode at densities below $\sim0.18\,\mathrm{cm}^{-3}$ the SN rate has dropped by a more than one order of magnitude. This highlights the complex interplay between SFR, clustering and feedback. In reality, however, all processes are at work as realised in model \textit{SWRC} (bottom right panel). 
	\label{fig:ambientsn}}
\end{figure*}

In Fig. \ref{fig:ambientsn}, we present the ambient ISM densities at the type II SN explosion sites for the six models. Those densities are computed as the average density of the gas cells within the SN injection radius (${r_\mathrm{inj} = 3 \times \Delta x}$ $\sim11.7\,\mathrm{pc}$). We want to point out again that we do not change the gas structure within those cells. Any density fluctuations before the injection of the thermal SN energy are retained (see Sec. \ref{sec:methods}). In the left panels, we show the runs without radiation and in the right panels the respective runs with radiation. The runs without radiation show clear bi-modal ambient density distributions with early SNe typically exploding at high densities similar to the star formation threshold $n_\mathrm{sink} = 10^3\,\mathrm{cm}^{-3}$. To highlight this, the blue dashed histograms indicate the ambient densities of the first two SNe in each cluster, which are typically high for \textit{S}, \textit{SW}, and \textit{SWC} at around $\sim20-200\,\mathrm{cm}^{-3}$. At such high densities, radiation losses are significant for the first SNe in each cluster. 
The strong clustering (e.g. 184 massive stars per cluster on average in simulation \textit{S}, (see Table \ref{tab:cluster}), however, less subsequent SNe explode in previously created bubbles, resulting in very low ambient densities creating the low-density peak. 

This is indicated by the grey shaded histograms, which include 80 per cent of all low-density SNe up to their limiting density of $n_\mathrm{ambient}^{80}$. This density is below $\sim10^{-2}\,\mathrm{cm}^{-3}$ for all simulations without radiation. The SN-only run \textit{S} has by far the highest SNR, as well as the broadest distribution with a double-peaked shape. The total number of SNe gets reduced by a factor of $\sim3.7$ by the stellar winds (models \textit{SW} and \textit{SWC}). The overall shape of the distribution is, however, very similar to model \textit{S} and the first SNe still explode only at the highest densities. Adding cosmic rays (\textit{SWC}) does not change this feature.

The ambient density distribution changes qualitatively with the inclusion of radiation (\textit{SR, SWR, SWRC}, in the right panels of Fig. \ref{fig:ambientsn}). The bi-modal nature disappears and due to the creation of lower density HII regions already the first SNe can explode in much lower ambient density environments. This is highlighted with blue dashed histograms in the right panels of Fig. \ref{fig:ambientsn}. One might assume that the early creation of HII regions results in even lower density for subsequent supernova explosions. This, however, is not the case as for all radiation models the cluster masses and number of massive star per cluster is significantly reduced. For example, the radiation run \textit{SR} has $\sim15$ massive stars per cluster compared to 184 massive stars per cluster in simulation \textit{S} (see Table \ref{tab:cluster}). As a result of this strongly reduced clustering less SNe explode in previously created bubbles, which - somewhat counter-intuitively - increases $n_\mathrm{ambient}^{80}$ by about one order of magnitude compared to the respective simulation without radiation \citep[see][for similar trends in a high-resolution dwarf galaxy simulation]{Hu2017}. The fraction SNe realised with only momentum injection (unresolved Sedov blast waves at high ambient densities) is below $\sim7.5$ per cent for all simulations and never drops below ${4.0\,\mathrm{per}\,\mathrm{cent}}$ (\textit{SWR}, see Table \ref{tab:cluster}). 

\section{ISM structure}\label{sec:structure}

\begin{figure}
	\centering
	\includegraphics[width=.99\linewidth]{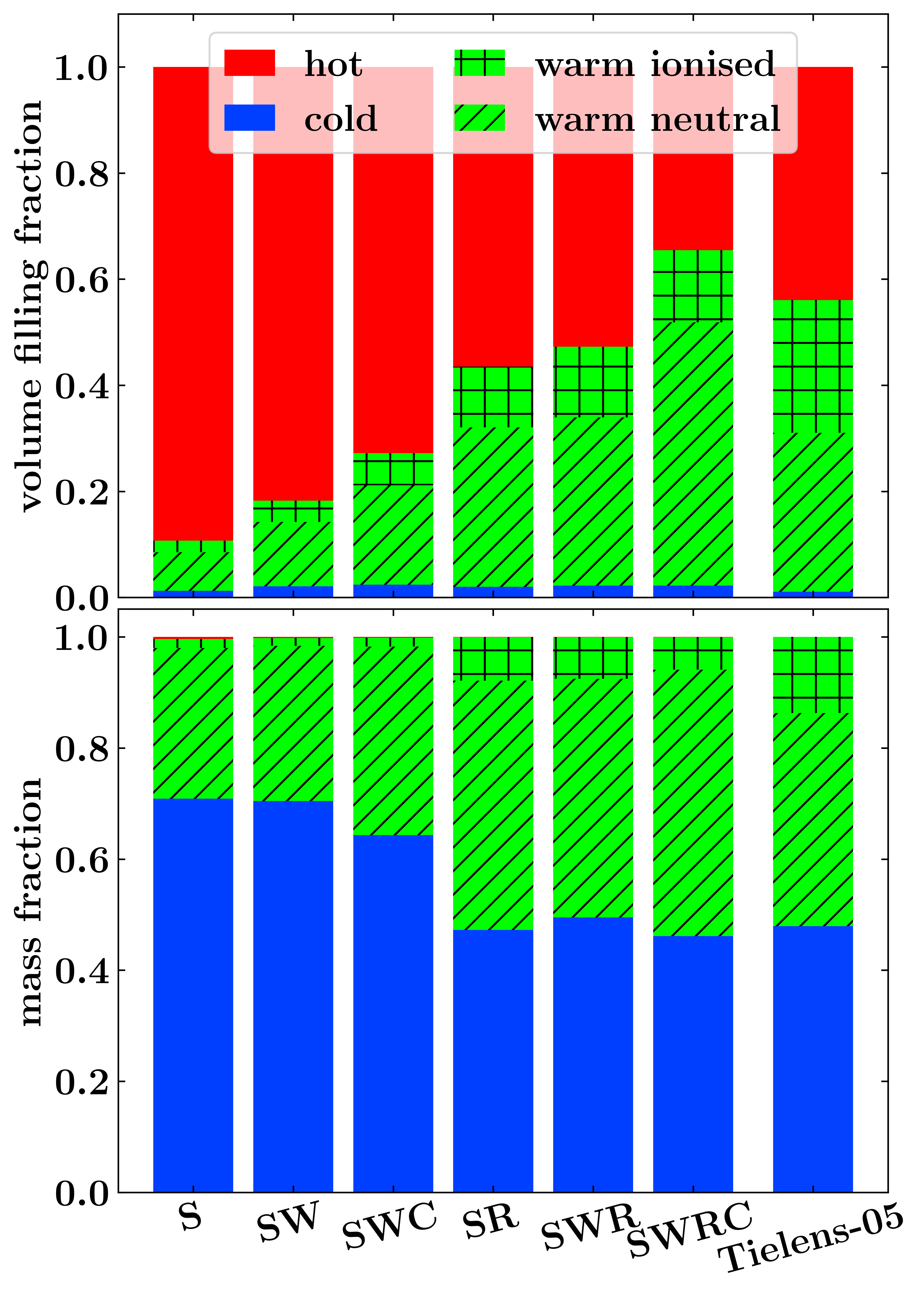}
	\caption{Volume-filling factors (VFF, top panel) and mass fractions (MF, bottom panel) of the mid-plane ISM ($z = \pm 250\,\mathrm{pc}$) for the cold phase (${T_\mathrm{cold} \leq 3 \times 10^{2}\,\mathrm{K}}$), the warm phase ($3 \times 10^{2}\,\mathrm{K}<$ $T_\mathrm{warm} \leq$ $3 \times 10^{5}$ K), and the hot phase ($T_\mathrm{hot} > 3 \times 10^{5}$ K) for all six simulations. The values are time averages from $t = 25-100\,\mathrm{ Myr}$. The warm gas phase is separated into neutral gas (diagonal shading) and ionised gas (chequered). Observational estimates for the solar neighbourhood ISM are given in the last columns \citep{Tielens2005}. The simulation \textit{S} with only SN feedback has the highest hot gas volume-filling factor $\sim90$ per cent. The inclusion of radiation strongly increases the VFF of the warm gas phase, also by adding photo-ionised gas, at the expense of the hot gas VFF. The inclusion of CRs further increases the warm gas VFF slightly. The cold gas MFs in runs including radiation are lower by $\sim10-20$ per cent compared to their non-radiation counterparts and agree better with observational estimates. \label{fig:vffmf}}
\end{figure}

In Fig. \ref{fig:vffmf}, we compare the volume-filling factors (VFFs) and mass fractions (MFs) of the mid-plane ISM of our simulations within $z = \pm 250$ pc. We separate the gas into a cold phase ($T \leq 300$ K), a warm phase (300 K $<$ $T \leq 3 \times 10^5$ K), and a hot phase ($T > 3 \times 10^5$ K). The warm phase we further divide into neutral and ionised gas based on the ionisation degree in the simulation cells. The last column in each panel give observational estimates for the solar neighbourhood ISM as stated in \citet{Tielens2005}. The VFFs and MFs are temporally averaged over $t = 25-100$ Myr. All values are listed in Table \ref{tab:mfvff}. 

Overlapping SN remnants are the main agent for creating the hot gas phase and therefore the volume of the simulation with only SNe (\textit{S}) is dominated by hot gas. The addition of stellar winds and CRs increases the VFF of warm gas up to $\sim25$ per cent, with a small ($\sim6$ per cent) contribution of collisionally ionised warm gas. Ionising radiation has the strongest impact on the mid-plane ISM. Resulting from recombined gas which was ionised in HII regions, the warm gas volume-filling factors increase by a factor $\sim3$ compared to the respective simulations without radiation. The volume of the hot phase is reduced correspondingly. This leaves the hot gas with a VFF of about 35-55 per cent which is also in line with observations and observation-based models (\citealt{Ferriere2001, Kalberla2009} and references therein). Estimates of the cold gas volume-filling factors (VFF$_\mathrm{cold}$) are more controversial, ranging from VFF$_\mathrm{cold} \sim5-18$ per cent for the mid-plane ISM \citep{Kalberla2009} down to VFF$_\mathrm{cold} \sim1$ per cent \citep{Tielens2005}.

Ionising UV radiation has also the strongest impact on the mass fractions of the ISM (MF, the lower panel of Fig. \ref{fig:vffmf}). The total warm gas MF increases from $\sim30-35$ per cent in non-radiation runs to $\sim50-55$ per cent in the radiation runs \textit{SR}, \textit{SWR}, \textit{SWRC}. The warm ionised MF increases by a factor $\sim4-5$ and the warm neutral gas MF by a factor $\sim1.5$. The gas reservoir for the increased mass of warm gas is the cold phase, which gets heated by the introduction of radiative feedback from massive stars. The cold mass fraction therefore decreases from $\sim70$ per cent to $\sim50$ per cent.
CRs slightly decrease the cold gas MF by $\sim5$ percentage points in \textit{SW} down to $\sim65$ per cent in \textit{SWC}. A similar, albeit weaker, trend can also be seen in the comparison of \textit{SWR} and \textit{SWRC}. The additional CR pressure smooths out the gas (very clearly seen in the outflow structure in Fig. \ref{fig:evol_2}, left panels), which prevents - to some extent - the creation of cold gas clumps. Instead, the gas is kept in a warm, diffuse regime.
Stellar winds do not impact the mass budget of the ISM. The addition of early wind feedback does not significantly change the MFs between \textit{S} and \textit{SW}, or \textit{SR} and \textit{SWR}. 
The mass in the hot phase is negligible in all models, as well as in observationally motivated estimates.
Our models including radiation (\textit{SR}, \textit{SWR}, \textit{SWRC}) all agree very well with observations, whereas the models without radiation (\textit{S}, \textit{SW}, \textit{SWC}) over-estimate the cold gas MFs, while under-estimating the warm ionised MFs.

\begin{table}
    \caption{Time-averaged ($t = 25-100\,\mathrm{Myr}$) mid-plane ISM ($z = \pm 250\,\mathrm{pc}$) volume-filling factors and mass fractions of the cold: $T_\mathrm{cold} \leq 3 \times 10^{2}$ K, the warm: $3 \times 10^{2}$ K $<$ $T_\mathrm{warm} \leq$ $3 \times 10^{5}$ K, and the hot: $T_\mathrm{hot} > 3 \times 10^{5}$ K phase. The warm phase is split into ionised and neutral gas. of. The given uncertainties include 1$\sigma$. Observational estimates for the local ISM are taken from \citet{Tielens2005}. The simulations are indicated in the first columns.}
    \begin{tabular}{lcccc}
        \hline
        Run & $\overline{\mathrm{VFF}}_\mathrm{hot}$ & $\overline{\mathrm{VFF}}_\mathrm{warm}^\mathrm{ionised}$ & $\overline{\mathrm{VFF}}_\mathrm{warm}^\mathrm{neutral}$ & $\overline{\mathrm{VFF}}_\mathrm{cold}$\\
          & [\%] & [\%] & [\%] & [\%]\\
        \hline
        \textit{S} & 89 $\pm$ 10 & 2 $\pm$ 2 & 7 $\pm$ 7 & 1 $\pm$ 1\\
        \textit{SW} & 82 $\pm$ 7 & 4 $\pm$ 2 & 12 $\pm$ 6 & 2 $\pm$ 1\\
        \textit{SWC} & 73 $\pm$ 6 & 6 $\pm$ 2 & 19 $\pm$ 5 & 2 $\pm$ 1\\
        \textit{SR} & 56 $\pm$ 16 & 11 $\pm$ 7 & 30 $\pm$ 10 & 2 $\pm$ 1\\
        \textit{SWR} & 53 $\pm$ 16 & 13 $\pm$ 9 & 32 $\pm$ 9 & 2 $\pm$ 1\\
        \textit{SWRC} & 35 $\pm$ 26 & 14 $\pm$ 7 & 50 $\pm$ 21 & 2 $\pm$ 1\\
        Tielens-05 & $\sim50$ & 25 & 30 & 1.05\\
        \hline
         Run & $\overline{\mathrm{MF}}_\mathrm{hot}$ & $\overline{\mathrm{MF}}_\mathrm{warm}^\mathrm{ionised}$ & $\overline{\mathrm{MF}}_\mathrm{warm}^\mathrm{neutral}$ & $\overline{\mathrm{MF}}_\mathrm{cold}$\\
          & [\%] & [\%] & [\%] & \\
        \hline
        \textit{S}    & 0.4 $\pm$ 0.3 & 2 $\pm$ 1 & 27 $\pm$ 7 & 71 $\pm$ 7 \\
        \textit{SW}   & 0.2 $\pm$ 0.1 & 1.4 $\pm$ 0.4 & 28 $\pm$ 6 & 70 $\pm$ 6 \\
        \textit{SWC}  & 0.2 $\pm$ 0.1 & 1.6 $\pm$ 0.5 & 34 $\pm$ 4 & 64 $\pm$ 4 \\
        \textit{SR}   & 0.05 $\pm$ 0.02 & 8 $\pm$ 5 & 45 $\pm$ 12 & 47 $\pm$ 14 \\
        \textit{SWR}  & 0.04 $\pm$ 0.01 & 8 $\pm$ 5 & 43 $\pm$ 8 & 50 $\pm$ 11 \\
        \textit{SWRC} & 0.03 $\pm$ 0.01 & 6 $\pm$ 3 & 48 $\pm$ 11 & 46 $\pm$ 13 \\
        Tielens-05 & - & 14 & 38 & 48 \\
        \hline
    \end{tabular}
    \label{tab:mfvff}
\end{table}

We list the average kinetic, thermal, magnetic, and CR energy densities of the mid-plane ISM in Table \ref{tab:edens} and compare to observational estimates summarised in \citet{Draine2011} and references therein.
Direct magnetic field strength measurements of the star-forming ISM via the Zeeman effect are only feasible in the dense ($n \gtrsim 10\,\mathrm{cm}^{-3}$) and cold neutral medium \citep[see e.g.][]{Heiles2005, Crutcher2019}. To better compare with observations, we therefore only average the magnetic energy densities over the atomic hydrogen gas below $T_\mathrm{cold} = 300\,\mathrm{K}$ in our mid-plane region. The kinetic, thermal and cosmic ray energy densities are volume-weighted over the full mid-plane ($|z| = 250\,\mathrm{pc}$).
Overall, the simulations including ionising UV radiation result in kinetic, thermal, and CR energy densities comparable to local neighbourhood ISM conditions. With only SNe (\textit{S}), the ISM is dominated by the hot phase with high-velocity gas, resulting in too high thermal and kinetic energy \citep[see][]{Walch2015}. Only when a warm gas phase is present, generated mostly by radiation, the energy densities become comparable to observations.
The CR energy densities in runs \textit{SWC} and \textit{SWRC} are within a factor of $\sim2$ close to the canonical local ISM value of $e_\mathrm{cr} = 1.39\,\mathrm{erg~cm}^{-3}$, supporting our model choices for the CR injection efficiency and the CR diffusion parameter.

\begin{table}
    \caption{Average mid-plane ($z = \pm 250$ pc) kinetic, thermal, magnetic, and CR energy densities $\overline{e}$ with 1$\sigma$ each for the six models for $t = 25-100\,\mathrm{Myr}$. To meaningful compare with observations, the magnetic energy densities are measured in cold ($T < 300\,\mathrm{K}$) neutral hydrogen gas (CNM), whereas the kinetic, thermal and CR energy densities are averaged over the entire mid-plane volume. Literature values - taken from \citet{Draine2011} and references therein - are estimates for solar neighbourhood ISM conditions. Without radiative feedback, especially thermal and kinetic energies are higher than observed.}
    \begin{tabular}{lcccc}
        \hline
         Run & $\overline{e}_\mathrm{kin}$ & $\overline{e}_\mathrm{th}$ & $\overline{e}_\mathrm{mag,CNM}$ & $\overline{e}_\mathrm{cr}$\\
          & [erg cm$^{-3}$] & [erg cm$^{-3}$] & [erg cm$^{-3}$] & [erg cm$^{-3}$]\\
        \hline
        \textit{S}    & 1.54 $\pm$ 1.31 & 3.36 $\pm$ 2.37 & 1.16 $\pm$ 0.97 & - \\
        \textit{SW}   & 1.10 $\pm$ 0.72 & 1.50 $\pm$ 0.75 & 0.91 $\pm$ 0.55 & - \\
        \textit{SWC}  & 0.91 $\pm$ 0.68 & 1.13 $\pm$ 0.75 & 0.80 $\pm$ 0.56 & 0.66 $\pm$ 0.37 \\
        \textit{SR}   & 0.49 $\pm$ 0.24 & 0.71 $\pm$ 0.21 & 1.05 $\pm$ 0.35 & - \\
        \textit{SWR}  & 0.42 $\pm$ 0.19 & 0.62 $\pm$ 0.23 & 1.06 $\pm$ 0.32 & - \\
        \textit{SWRC} & 0.34 $\pm$ 0.10 & 0.51 $\pm$ 0.28 & 0.86 $\pm$ 0.43 & 0.87 $\pm$ 0.63 \\
        Draine-10 & 0.22 & 0.49 & 0.89 & 1.39 \\
        \hline
    \end{tabular}
    \label{tab:edens}
\end{table}

\section{Implications for galactic outflows}\label{sec:outflow}

For investigating the outflow energetics, we define an energy loading $\gamma$ \citep[see e.g.][]{Kim2017} as
\begin{equation}
    \gamma = \frac{\dot{E}_\mathrm{out}}{\overline{\dot{E}}_\mathrm{inj}},
\end{equation} where $\dot{E}_\mathrm{out}$ is the outflowing energy rate measured at $z = \pm 1\,\mathrm{kpc}$, consisting of thermal, kinetic, magnetic and CR energy, and $\overline{\dot{E}}_\mathrm{inj}$ is the average energy injection rate into the ISM, consisting of energy injection from SNe, stellar winds, ionising UV radiation and CRs. 

Similarly, we define a mass loading $\eta$ as the ratio of the mass outflow rate, $\dot{M}_\mathrm{out}$ measured at $z = \pm 1\,\mathrm{kpc}$ divided by the time-averaged instantaneous SFR $\overline{\dot{M}}_\star$,
\begin{equation}
    \eta = \frac{\dot{M}_\mathrm{out}}{\overline{\dot{M}}_\star}.
\end{equation}

Finding a useful working definition for the above loading factors is slightly complicated \citep[see discussion in][]{Kim2017}. The energy injected into the mid-plane in one time-step does not instantaneously influence the energy outflow at a height of $z = \pm 1\,\mathrm{kpc}$, just as stars formed in the mid-plane do not correlate with the instantaneous mass outflow rate. One possibility is to introduce a time delay $\Delta t = \Delta z \times \Tilde{v}^{-1}$, with a characteristic speed $\Tilde{v}$ of the gas in the ISM. This would, however, assume that the gas flows funnel-like straight from the birth site of stars to the outflow region $z = \pm 1\,\mathrm{kpc}$. In reality, the gas is turbulent and the impact of local and temporal overlapping star formation events is non-linear. Another solution could be the use of moving averages but the choice of the window size is arbitrary and the resulting mean values can vary for more than 60 per cent compared to a global mean. We find the most robust definition is to take the ratio of the respective outflow rates and the global averaged SFR and energy injection rates, $\overline{\dot{M}}_\star$ and $\overline{\dot{E}}_\mathrm{inj}$. 

Quoted mean values for the energy loading and mass loading then are averaged over $t = 25-100\,\mathrm{Myr}$. To compare all models, we take the averages from the beginning of star formation (which is identical in all six runs), instead of the onset of an outflow. Therefore the averages are also taking into account episodes in which no or very weak outflows are present. This is the case for about $\sim10$ per cent of the time in \textit{S}, \textit{SW}, \textit{SWC}, $\sim65$ per cent in \textit{SR} and \textit{SWR}, and $\sim40$ per cent in \textit{SWRC}.
In Table \ref{tab:loadings}, we give an overview of the mean SFR surface density $\overline{\Sigma}_{\dot{M}_\star}$, the mean energy loading factors, normalised to SN injection energy $\overline{\gamma}_\mathrm{sn}$, the mean mass loading factors $\eta$, and their fractional compositions from our simulations.

\subsection{Energy loading}\label{sec:gamma}

\begin{figure}
	\centering
	\includegraphics[width=.99\linewidth]{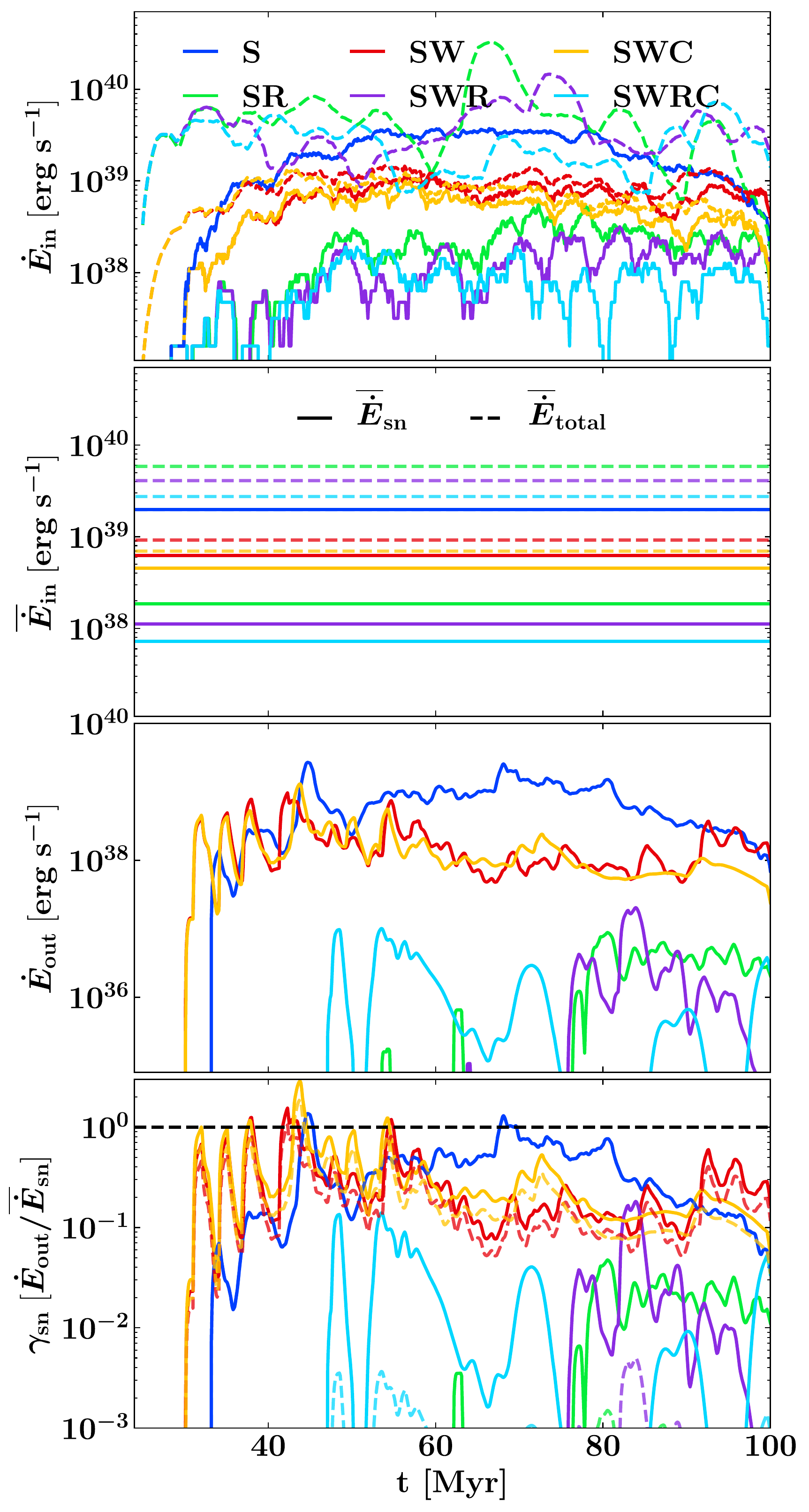}
	\caption{Top panel: SN energy injection rates (solid lines) and total energy injection rates (dashed lines) vs. time for the six models. The total energy injection rates include the wind luminosity (models \textit{W}), the UV luminosity (models \textit{R}), and the injected energy in CRs (models \textit{C}). 
	\nth{2} panel: Mean SN energy injection rates $\overline{\dot{E}}_\mathrm{sn}$ (solid lines) and average total energy injection rates $\overline{\dot{E}}_\mathrm{total}$ (dashed lines) for all models, derived from data shown in the top panel averaged over the time interval from 25 Myr to 100 Myr.
	\nth{3} panel: Total energy outflow rate measured at $z = \pm 1$ kpc vs. time. 
	Bottom panel: Energy loading $\gamma$, measured at $z = \pm 1$ kpc vs. time. The solid lines show $\gamma_\mathrm{sn}$, which are normalised to the average SN energy injection rate. The dashed lines show $\gamma_\mathrm{total}$, normalised to the averaged total energy injection rate. The dashed black line indicates an energy loading of unity. \label{fig:gamma}}
\end{figure}

In Fig. \ref{fig:gamma}, we present the energy rate budget of our simulations. In the top panel, we show the energy injected into the mid-plane ISM as a function of time. The solid lines show the energy injection rates of only SNe (not including the additional CR energy in the runs \textit{SWC} and \textit{SWRC}). The dashed lines indicate the total injected energy including winds, ionising UV radiation, and the CRs for the respective simulations. In the \nth{2} panel, we show the same quantities but averaged over ${t = 25-100\,\mathrm{Myr}}$, $\overline{\dot{E}}_\mathrm{total}$ and $\overline{\dot{E}}_\mathrm{sn}$. These values are used to compute the energy loading factors.
The integrated stellar wind and supernova energy injection rates are comparable, and the CR injection is 10 per cent of the SN rate by construction.
The total energy injected in the radiation runs is higher by $\sim2$ orders of magnitude. This is caused by the high UV photon luminosity, as expected from single stellar population models \citep[see e.g.][for a discussion about wind and UV luminosities]{Agertz2013, Peters2017}. However, in our simulations the injected radiation couples only weakly to the surrounding ISM and to the large-scale gas motions \citep{Peters2017}. For completeness, we present the energy injection by the different mechanisms in Appendix \ref{sec:inj}. In the \nth{3} panel, we show the total energy outflow rates measured at $z = \pm 1$ kpc, and in the bottom panel we show the energy loading $\gamma$, measured at $z = \pm 1$ kpc, normalised to the total injected energy (dashed lines) and normalised only to the injected SN energy (solid lines). The models with the highest SN energy injection rate (\textit{S, SW, SWC}) also have the highest energy outflow rate. Phases with SN energy loading values above unity result from the delayed impact of clustered SNe and the breakout of super-shells. Radiation couples inefficiently (runs \textit{SR, SWR, SWRC}) and the total energy loading values are about a factor of 30 lower than the respective no-radiation simulations.

In the following we only refer to average SN energy loading values $\overline{\gamma}_\mathrm{sn}$ (see Table \ref{tab:loadings}) as for the short time-scale simulations presented here SNe are the main driver for outflows. Also, these values can be better compared to the literature as most previous studies only include the SN feedback channel \citep[see e.g.][]{Fielding2017, Fielding2018, Kim2018, Li2017, Schneider2020}. In the SN-only run (\textit{S}), 34 per cent on average of the injected energy leaves the mid-plane (see Table \ref{tab:loadings}). This value decreases slightly to an average energy loading of $\gamma_\mathrm{sn} = 28-32$ per cent when accounting for stellar winds (\textit{SW}, \textit{SWC}). Model \textit{SWC} has a $\sim5$ percentage points higher energy loading than its counterpart without CRs (\textit{SW}), because the CR diffuse independent of the bulk gas motion out of the mid-plane ISM and carry most of their energy with them without significant cooling losses. The inclusion of radiation lowers the energy loading significantly to about 1 per cent due to the inefficient conversion of radiation energy to the gas kinetic energy \citep{Haid2018}. Even if all radiation energy was converted into kinetic energy, it might not result in a significant outflow since no hot gas will be generated. Ionising UV radiation only heats the gas to $T \approx 10^4\,\mathrm{K}$ generating velocities of about $v \approx 10-20 \,\mathrm{km s}^{-1}$. This is insufficient to overcome the external gravitational potential and lift the gas to heights of $z = 1\,\mathrm{kpc}$.

The out-flowing energy is initially dominated by thermal energy in run \textit{S} and becomes comparable to the kinetic energy in the later phases of simulation. A similar behaviour is seen in model \textit{SW}. The energy flux in all radiation runs is dominated by thermal energy. The situation qualitatively changes for runs with CRs. Here the energy flux is dominated by CR energy (see in particular simulation \textit{SWC}). The time evolution of the kinetic, thermal and CR energy flux normalised to the injected SN energy is shown in Fig. \ref{fig:gamma_i} in Appendix \ref{sec:phase_outflow} and the respective fractions of the average SN energy loadings are summarised in Table \ref{tab:loadings}. If thermal phases are considered (see Fig. \ref{fig:gamma_phase} in the Appendix and Table \ref{tab:loadings}), the energy flux of most simulations is dominated by hot gas. At later times, the energy loading in hot and warm gas become comparable for simulations \textit{S} and \textit{SW}. Adding CRs (run \textit{SWC}) shifts the budget towards warm gas in agreement with previous findings that CRs result in cooler and smoother outflows \citep{Girichidis2018}.

\begin{figure}
	\centering
	\includegraphics[width=.99\linewidth]{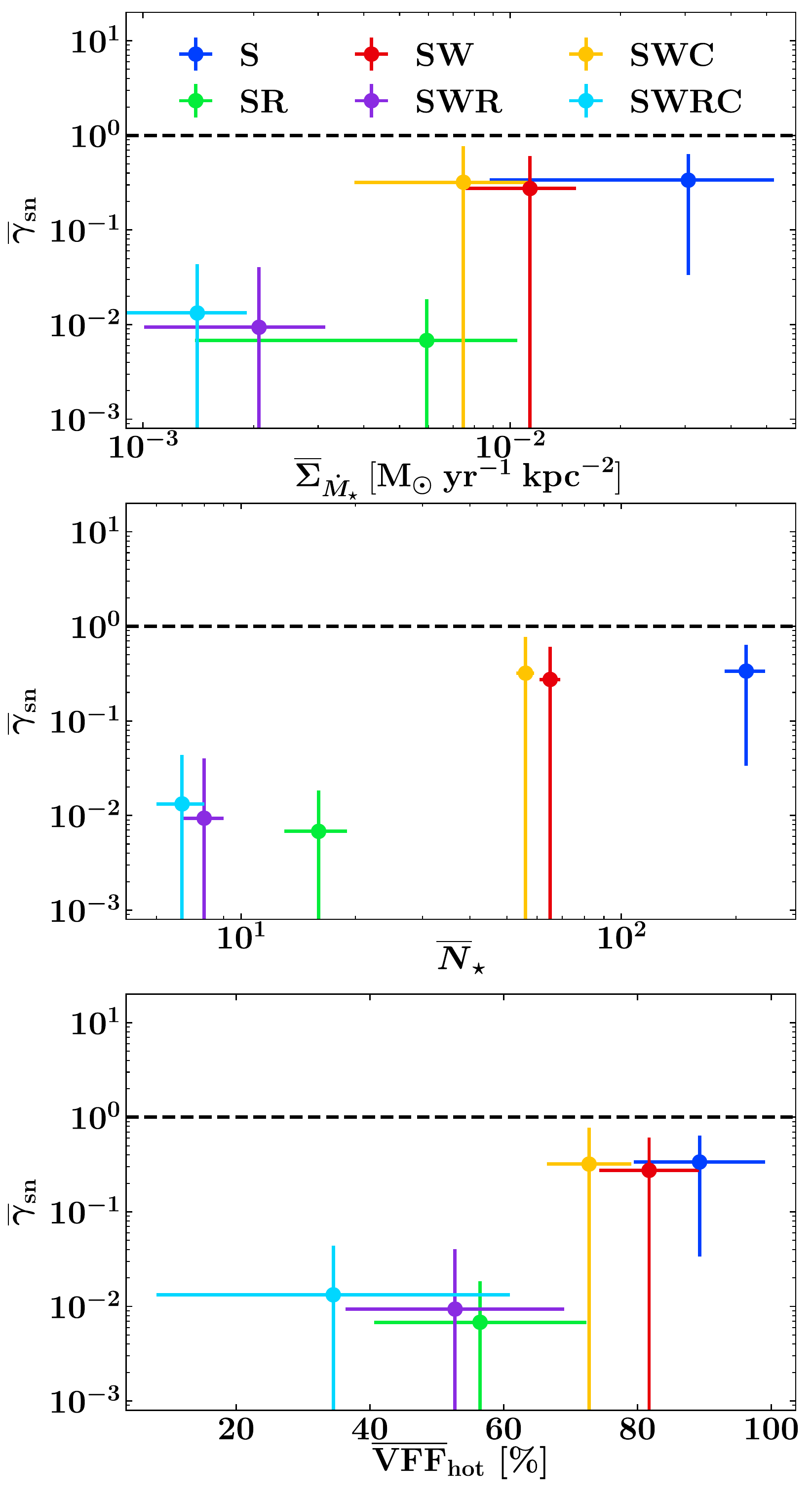}
	\caption{Average SN energy loading $\overline{\gamma}_\mathrm{sn}$ as a function of average SFR surface density (top panel), average number of massive stars per cluster (middle panel) and average hot gas VFF (bottom panel). Each value is given with 1$\sigma$ scatter. The energy loading correlates with SFR, stellar clustering, and hot gas fraction. We therefore expect typically low energy loading values at disk surface densities $\sim10 M_\odot\,\mathrm{pc}^{-2}$ for our simulation with the most complete set of physical models \textit{SWRC}. \label{fig:gamma_relations}}
\end{figure}

In Fig. \ref{fig:gamma_relations} we show the average SN energy loading factors $\overline{\gamma}_\mathrm{sn}$ as a function of the averaged SFR surface density $\overline{\Sigma}_{\dot{M}_\star}$ (top panel), of the average number of massive stars per cluster, the clustering, $\overline{N}_\star$, and of the averaged hot gas volume-filling factor $\overline{\mathrm{VFF}}_\mathrm{hot}$. Error-bars indicate a 1$\sigma$ standard deviation.
Simulations with the highest SFRs also have the highest energy loading factors (top panel of Fig. \ref{fig:gamma_relations}). The energy loading of \textit{S}, \textit{SW}, and \textit{SWC} is very similar, reflecting their comparable ambient SN density distributions and high volume-filling factors of the hot phase (see Fig. \ref{fig:ambientsn} and Fig. \ref{fig:vffmf}).

In the middle panel of Fig. \ref{fig:gamma_relations}, we see that the SN energy loading also correlates with stellar clustering. The energy loading is highest for the SN only model \textit{S}, which has the strongest clustering. Winds (\textit{SW} and \textit{SWC}) reduce the number of massive stars per cluster and $\overline{\gamma}_\mathrm{sn}$ is a factor of $\sim1.5-3$ lower. Radiation (\textit{SR}, \textit{SWR} and \textit{SWRC}) furthermore reduces the clustering of massive stars resulting in even lower energy loadings around $\sim1$ per cent. This analysis indicates that the clustering of massive stars and therefore of the SNe is an important agent for driving efficient outflows \citep[for similar conclusions see e.g.][]{Smith2020}.

We show the average energy loading $\overline{\gamma}_\mathrm{sn}$ as a function of the average hot gas volume-filling factor $\overline{\mathrm{VFF}}_\mathrm{hot}$ for the six models in the bottom panel of Fig. \ref{fig:gamma_relations}. As discussed above, ionising UV radiation decreases the SFR and the clustering of massive stars. Therefore, the SNR also decreases and fewer SN remnants overlap (see Sec. \ref{sec:sn}. This limits the creation of a volume-filling hot phase (see Sec. \ref{sec:structure}). Without additional driving mechanisms, the hot phase is the main agent for accelerating gas out of the mid-plane. On these short time-scales, the outflow driving by CRs has not yet set in. 

\subsection{Mass loading}\label{sec:eta}

\begin{figure}
	\centering
	\includegraphics[width=.99\linewidth]{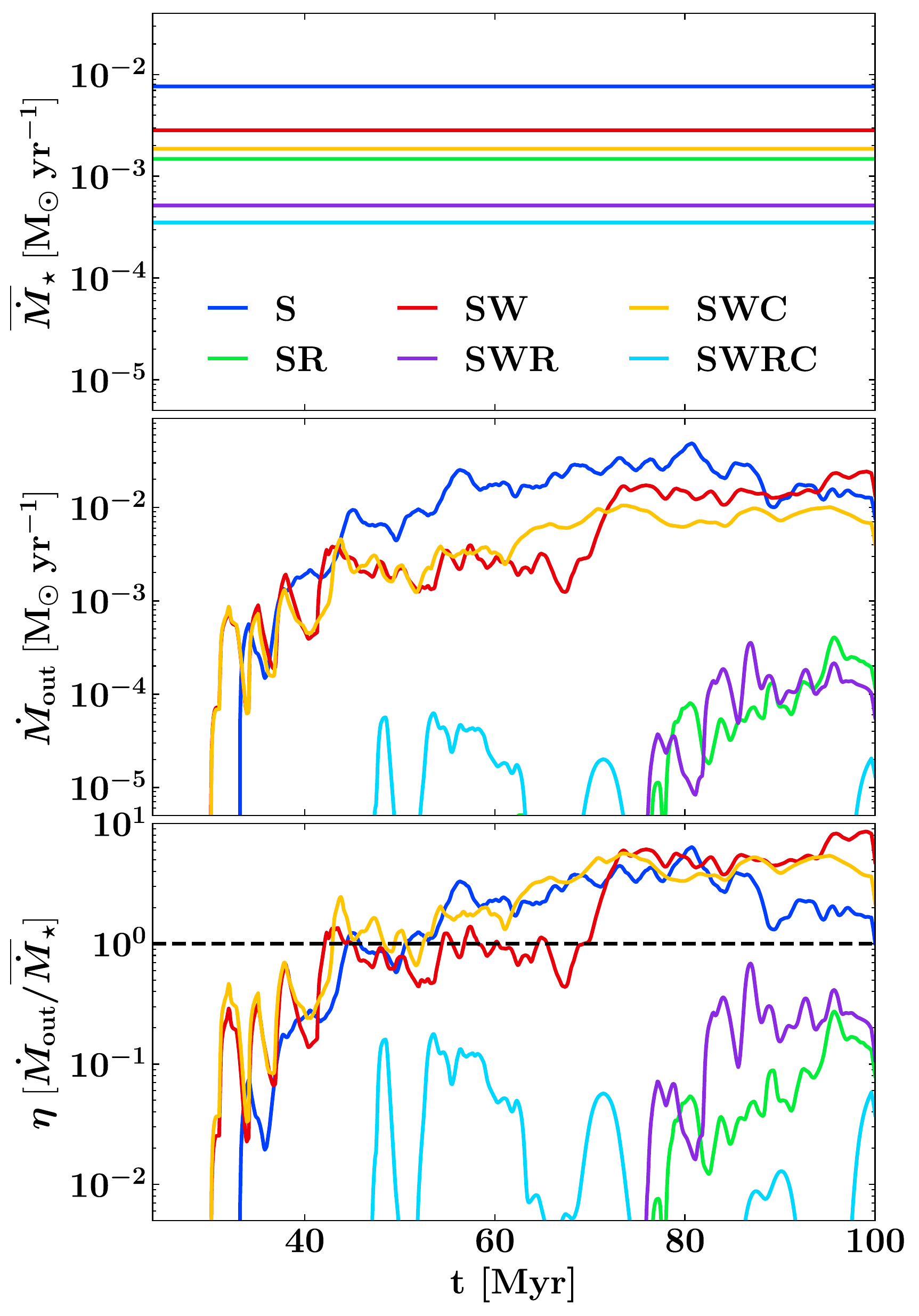}
	\caption{Average star formation rates $\overline{\Sigma}_\mathrm{\dot{M}_\star}$ (top panel), instantaneous mass outflow rates $\dot{M}_\mathrm{out}$ through $z = \pm 1\,\mathrm{kpc}$ (middle panel) and mass loading factors $\eta = \dot{M}_\mathrm{out} / \overline{\dot{M}}_\star$ at $z = \pm 1$ kpc (bottom panel) for the six simulations. Models with the highest SFRs also have the highest mass outflow rats and mass loading factors (\textit{S}, \textit{SW}, \textit{SWC}). Simulations with radiation, including our most complete model (\textit{SWRC}), do not drive strong outflows. CRs only become relevant for outflow driving on longer time-scales not presented in this study \citep[see e.g.][]{Girichidis2016, Girichidis2018}. \label{fig:eta}}
\end{figure}

In Fig. \ref{fig:eta}, we show the average star formation rates $\overline{\dot{M}}_\star$ (top panel), mass outflow rates through $z = \pm 1$ kpc (middle panel), and mass loading factors $\eta$ (bottom panel) for the six models from $t = 25$ Myr to $t = 100$ Myr. Qualitatively, the behaviour of the mass loading is similar to the energy loading discussed above.

\begin{figure}
	\centering
	\includegraphics[width=.99\linewidth]{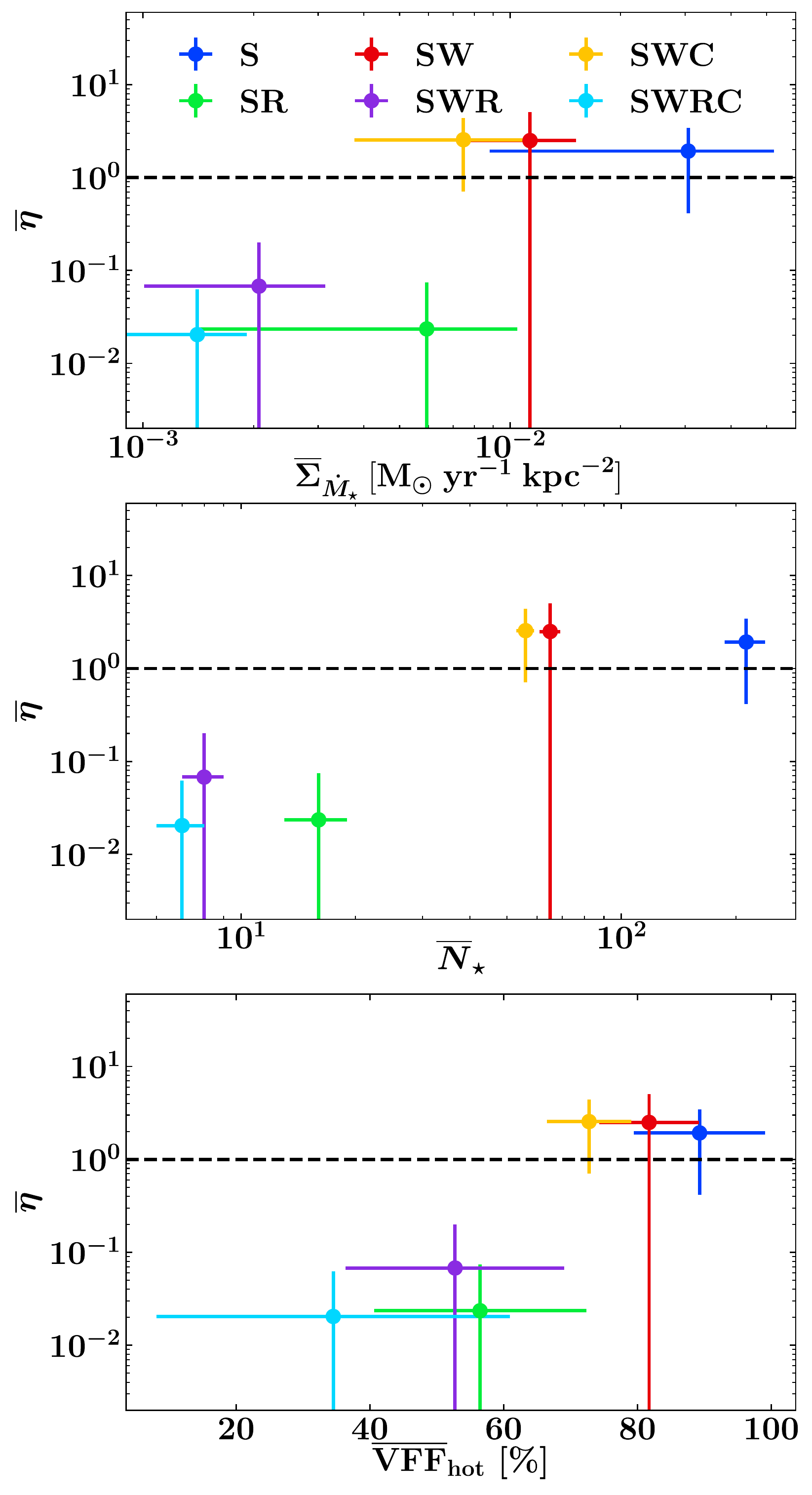}
	\caption{Average mass loading $\overline{\eta}$ as a function of SFR surface density (top panel), the average number of massive stars per cluster (middle panel) and averaged hot gas VFF (bottom panel). Each value is given with a 1$\sigma$ scatter. Similar to the energy loading, the mass loading positively correlates with all three properties. \label{fig:eta_relations}}
\end{figure}

Strong outflows are constantly driven by the hot phase generated by clustered SNe (compare with the edge-on view of the gas surface density in Fig. \ref{fig:evol_0} and also Fig. \ref{fig:vffmf}). For models with radiation (\textit{SR}, \textit{SWR}, \textit{SWRC}), the outflow is delayed and the outflow rates are at least one order of magnitude lower due to the lower SFRs, weaker clustering, and correspondingly lower hot volume-filling factors. The trends of increasing average mass loading with increasing SFR, the increasing average number of massive stars in clusters and average hot gas volume-filling factors are summarised in Fig. \ref{fig:eta_relations}. The interpretation of the trends is the same as for the energy loading discussed in the previous section. For the short time-scales after the onset of star formation presented in this study, SNe and their clustering are the main drivers for generating the hot phase which is powering the outflows \citep[see e.g.][]{Martin2012, Newman2012, Li2020}. For all simulations, the mass outflow is dominated by hot gas initially with the warm gas taking over soon after the onset of star formation in all simulations. This effect is strongest for \textit{SWC} (see Fig. \ref{fig:eta_phase} in the Appendix and Table \ref{tab:loadings}). This is also the only simulation with a measurable outflow in the cold phase at a very low fraction of 0.3 per cent (see Table \ref{tab:loadings}).

The effects of CRs are intricate. They do not directly impact star formation but can have a long-term influence on the outflow. Between \textit{SW} and \textit{SWC} there is not much difference in $\eta$ because their outflows are driven mostly by the hot phase generated in the mid-plane ISM, which has a comparable VFFs (see Sec. \ref{sec:structure}). The average hot gas VFF of \textit{SWRC}, on the other hand, is a factor of $\sim2$ lower which would result in a weaker outflow but the additional pressure gradient of the CRs helps to lift the gas away from the mid-plane, in alignment with observed mass loading factors $\eta$ of unity and above. On a time-scale for up to 100 Myr CR do not increase the driving of an outflow. However, previous idealised studies without self-consistent star formation indicate that with a longer evolution the additional CR pressure gradient becomes the dominant mechanism of driving outflows \citep{Simpson2016, Girichidis2016, Girichidis2018}. We will investigate this further in a follow-up study in which we focus on the long-term ($t = 300\,\mathrm{Myr}$) evolution of models \textit{SWR} and \textit{SWRC}, among others (Rathjen et al., in prep.).

\begin{table*}
    \caption{Averaged SFR surface densities $\overline{\Sigma}_{\dot{M}_\star}$, averaged SN energy loading $\overline{\gamma}_\mathrm{sn}$ normalised to only SN injection, and averaged mass loading $\eta$ with 1$\sigma$ each for the six models, and the fractions of the energy- and mass loading factors divided into their compositions and thermal phases to the averaged total energy- and mass loading factors, $f_{\gamma_i} = \overline{\gamma_i} / \overline{\gamma_\mathrm{sn}}$ and $f_{\eta_i} = \overline{\eta_i} / \overline{\eta}$. The indices kin, th and cr denote the energy split into kinetic, thermal and CR energy, respectively, whereas hot, warm and cold divide the gas into thermal phases as described in Sec. \ref{sec:structure}. The time evolutions are presented in the Appendix, Fig. \ref{fig:gamma_i} to Fig. \ref{fig:eta_phase}.}
    \begin{tabular}{lccccccccccccccc}
        \hline
        Run & $\overline{\Sigma}_{\dot{M}_\star}$ & $\overline{\gamma}_\mathrm{sn}$ & $\overline{\eta}$ & & $f_{{\gamma}_\mathrm{kin}}$ & $f_{\gamma_\mathrm{th}}$ & $f_{\gamma_\mathrm{cr}}$ & & $f_{\gamma_\mathrm{hot}}$ & $f_{\gamma_\mathrm{warm}}$ & $f_{\gamma_\mathrm{cold}}$ & & $f_{\eta_\mathrm{hot}}$ & $f_{\eta_\mathrm{warm}}$ & $f_{\eta_\mathrm{cold}}$ \\
         & [$\mathrm{M}_\odot$ yr$^{-1}$ kpc$^{-2}$] & [\%] &  & & [\%] & [\%] & [\%] & &  [\%] & [\%] & [\%] & & [\%] & [\%] & [\%] \\
        \hline
        \textit{S}    & (3.1 $\pm$ 2.2) $\times 10^{-2}$ & 33.6 $\pm$ 30.3 & 1.93 $\pm$ 1.51 & & 34.8 & 65.1 &    - & & 95.6 &  4.4 & 0.0 & & 29.0 & 71.0 & 0.0 \\
        \textit{SW}   & (1.1 $\pm$ 0.4) $\times 10^{-2}$ & 27.5 $\pm$ 33.4 & 2.51 $\pm$ 2.55 & & 31.3 & 68.5 &    - & & 96.1 &  3.9 & 0.0 & & 18.2 & 81.8 & 0.0 \\
        \textit{SWC}  & (7.5 $\pm$ 3.7) $\times 10^{-3}$ & 32.0 $\pm$ 45.1 & 2.55 $\pm$ 1.85 & & 16.0 & 47.5 & 36.4 & & 82.6 & 17.3 & 0.2 & & 14.5 & 85.2 & 0.3 \\
        \textit{SR}   & (5.9 $\pm$ 4.5) $\times 10^{-3}$ &  0.7 $\pm$  1.2 & 0.02 $\pm$ 0.05 & &  6.6 & 93.2 &    - & & 90.8 &  9.2 & 0.0 & & 45.5 & 54.5 & 0.0 \\
        \textit{SWR}  & (2.1 $\pm$ 1.1) $\times 10^{-3}$ &  0.9 $\pm$  3.1 & 0.07 $\pm$ 0.13 & &  9.9 & 89.9 &    - & & 98.1 &  1.9 & 0.0 & & 35.0 & 65.0 & 0.0 \\
        \textit{SWRC} & (1.4 $\pm$ 0.5) $\times 10^{-3}$ &  1.3 $\pm$  3.0 & 0.02 $\pm$ 0.04 & &  1.5 & 90.2 &  8.2 & & 99.9 &  0.1 & 0.0 & & 93.7 &  6.3 & 0.0 \\
        \hline
    \end{tabular}
    \label{tab:loadings}
\end{table*}

\begin{figure}
	\centering
	\includegraphics[width=.99\linewidth]{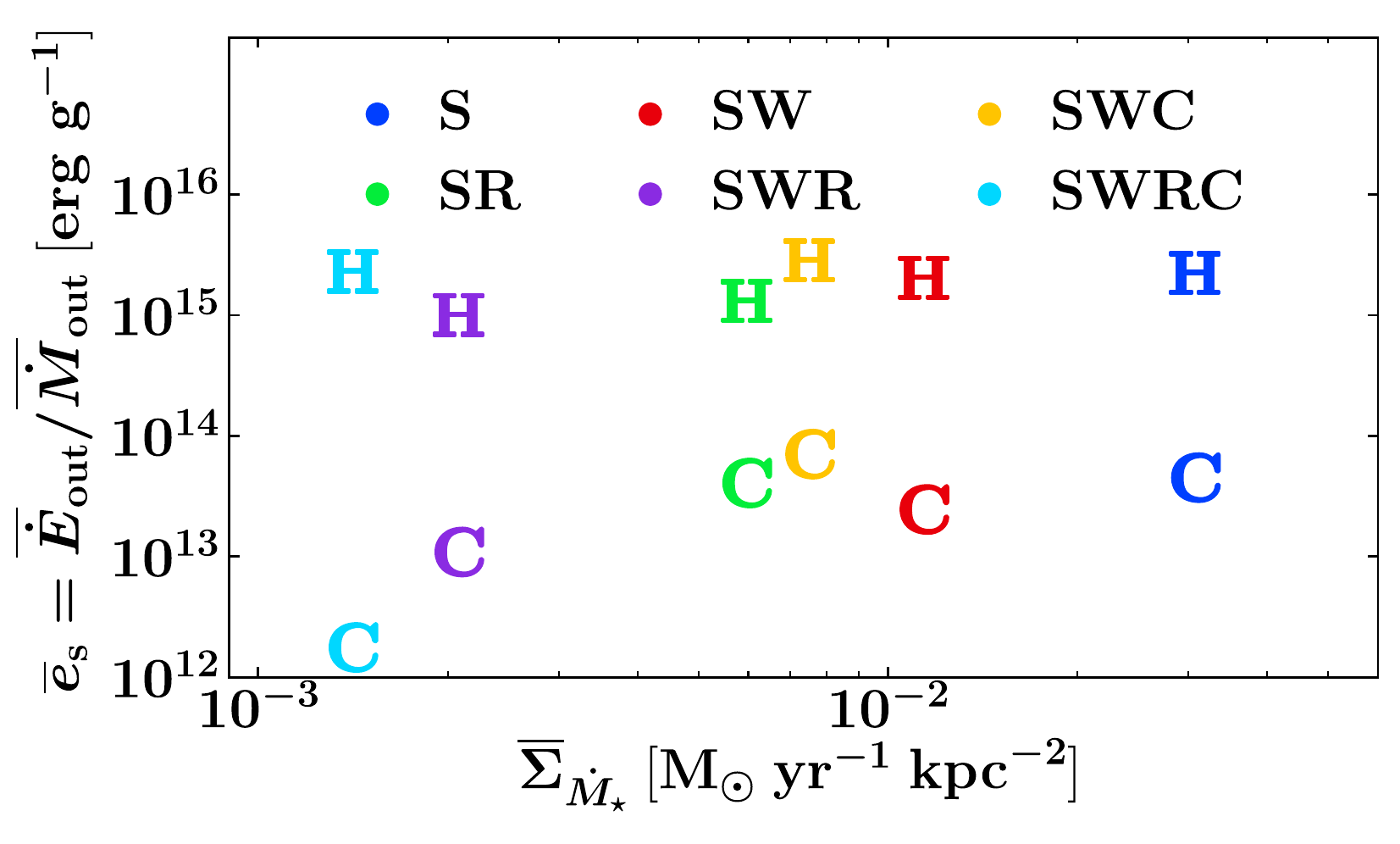}
	\caption{Averaged specific energy $\overline{e}_\mathrm{s}$ of the hot and cool gas outflow as a function of $\overline{\Sigma}_{\dot{M}_\star}$. Here, we define the cool gas as the combination of our cold and warm phase defined in Sec. \ref{sec:structure}. The specific energy of the hot outflow in our six models is independent of the SFR and the stellar feedback processes.\label{fig:e_spec}}
\end{figure}

In Fig. \ref{fig:e_spec} we show the average specific energy of the hot and cool gas outflow $\overline{e}_\mathrm{s} = \overline{\dot{E}}_\mathrm{out} / \overline{\dot{M}}_\mathrm{out}$ as a function of the average SFR surface density $\overline{\Sigma}_{\dot{M}_\star}$ for the six models. Here, we define the cool gas phase as the sum of the cold and warm gas as defined in Sec. \ref{sec:structure} ($T_\mathrm{cool} \leq 3 \times 10^5\,\mathrm{K}$, $T_\mathrm{hot} > 3 \times 10^5\,\mathrm{K}$). Even though the total mass and energy outflow rates, as well as the composition of the outflows, vary strongly between the different models (Fig. \ref{fig:gamma}, \nth{3} panel and Fig. \ref{fig:eta}, \nth{2} panel), the specific energy of the hot gas outflows only varies by a factor $\sim2$. The specific energy of the cool outflow, however, varies by a factor $\sim20$. \citet{Li2020} compiled a list of the outflow properties in recent small-box ISM simulations with only SN feedback (for a detailed list of considered simulations see their Table 1). Note that the definition of the thermal phases and the height in which the outflows are measured differ slightly in each work but are broadly comparable to ours. \citet{Li2020} find that the hot gas outflow specific energy $e_\mathrm{s,\,hot}$ only varies within a factor 30 ($e_\mathrm{s,\,hot} \approx 3.16 \times 10^{14}-10^{16}\,\mathrm{erg}\,\mathrm{g}^{-1}$), while the SFR surface densities in their examined simulations vary over 4 orders of magnitude between $\Sigma_\mathrm{\dot{M}_\star} = 10^{-4}-1\,\mathrm{M}_\odot\,\mathrm{yr}^{-1}\,\mathrm{kpc}^{-2}$, much in agreement with our results. However, they report a large spread for the ratio of the hot and cool outflow specific energy of $e_\mathrm{s,\,hot} / e_\mathrm{s,\,cool} \approx 10-1000$, whereas we find a lower ratio of $e_\mathrm{s,\,hot} / e_\mathrm{s,\,cool} \approx 50$ for the SN-only model \textit{S}. In all our models the hot gas outflows have higher specific energy than the cool gas outflows and therefore can travel further away from the mid-plane ISM and have a possibly larger impact on the CGM.

\section{Discussion}\label{sec:discussion}

Although many recent works studied the solar neighbourhood ISM with its outflow properties, direct comparison is not always feasible, since most studies are omitting some major stellar feedback processes - most notably ionising UV radiation from massive stars and self-gravity - included in our framework. In the following, we will try to contextualise our results by comparing them to some of those recent studies. This comparison, however, is by no means exhaustive.

\citet{Butler2017} simulate a self-gravitating kilo-parsec region of a galactic disc with self-consistent star formation and feedback in form of SN and dissociating and ionising UV radiation, down to a resolution of $\Delta x$ = 0.5 pc. Their setup is inherited from a shearing, global disc simulation with a flat rotation curve. The kpc-sized patch sits at a galactocentric radius of 4.25 kpc with a ${\Sigma_\mathrm{gas} = 17 \mathrm{M}_\odot\,\mathrm{pc}^{-2}}$. The boundary conditions, however, are not of a shearing box. Therefore, they only simulate for 20 Myr, i.e. the flow crossing time of the maximum shear velocity in their setup. They argue that those short time-scales are enough to reach quasi-statistical equilibrium conditions \citep[for a similar conclusion see e.g.][]{Pellegrini2020}. SN feedback is realised by injecting 10$^{51}$ erg of kinetic energy $3\,\mathrm{Myr}$ after the birth of an individual massive star. They model H$_2$ dissociation and photo-ionisation of hydrogen via radiative transfer with a reduced speed of light approximation. They achieve SFR surface densities in agreement with observational data from \citet[][same data-set as we use for compression in Sec. \ref{sec:sfr}]{Bigiel2008} with a model including both radiation types and SN. In their analysis, they pick two regions (patches with $400 \times 400\,\mathrm{pc}^2$ and $400 \times 600\,\mathrm{pc}^2$), comparable in size to our mid-plane ISM definition. Only accounting for SN leads to very high SFR surface densities ($\sim5-50\,\mathrm{M}_\odot\,\mathrm{yr}^{-1}\,\mathrm{kpc}^{-2}$) in those regions, time-averaged from 5 to 10 Myr. However, spatially averaged on a kilo-parsec scale and time-averaged from 15 to 20 Myr, SN alone can already lead to reasonable SFRs in their models. They argue - similar to us - that the star-burst resulting from the lack of early feedback leads to a strong expulsion of gas out of the star-forming regions, which then leads to a regulation of star formation.
Radiation is needed to properly model the chemical state and temperature structure of the ISM. Their reported mass fractions of ionised hydrogen is with $\sim10-15$ per cent comparable to the warm ionised gas mass fraction in our \textit{SR} model of $\overline{\mathrm{MF}}^\mathrm{ionised}_\mathrm{warm} = (8 \pm 5)$ per cent (see Table \ref{tab:mfvff}).

\citet{Dobbs2011} simulate the star-forming ISM in disc galaxies including a galactic potential, heating from the ISRF, cooling, self-gravity, H$_2$ chemistry and SN feedback. They do not explicitly include star formation via a sub-grid sink particle approach but instead track when a pocket of gas fulfils the criteria for star formation. When this is the case, they realise SNe by injecting thermal energy with $10^{51}\,\mathrm{erg}$ per $160\,\mathrm{M}_\odot$ of new stars that would have been formed times a star formation efficiency (SFE) $\epsilon$ into the medium. Those star formation criteria are very similar to ours (see Sec. \ref{sec:methods}): gas density has to be greater than $10^3$ particles per $\mathrm{cm}^{-1}$, the gas flow has to be converging, and it has to be gravitationally bound.
With this prescription, they obtain reasonable ISM conditions and SFRs along the Kennicutt-Schmidt relation with only SN feedback but only for star formation efficiencies of $\epsilon \approx 0.05-0.2$.

Similarly, \citet{Tress2020} simulate an interacting M51-like galaxy with self-gravity, a non-equilibrium, time-dependent chemical network, self-consistent star formation with sink particles and stellar feedback only in the form of SNe. Also, their sink particle formation prescription and accretion parameters are nearly identical to ours with the criteria mentioned above, with the big difference that they, too, impose an artificial star formation efficiency of $\epsilon = 0.05$. SNe then are realised similarly as we do with thermal energy injection of $10^{51}\,\mathrm{erg}$ and momentum injection, if the Sedov-Taylor phase is unresolved to prevent over-cooling. Again, this study can reproduce an ISM within observable scatter and a total SFR of $4\,\mathrm{M}_\odot\,\mathrm{yr}^{-1}$, which is very close to the observed value for M51 of $4.6\,\mathrm{M}_\odot\,\mathrm{yr}^{-1}$ \citep{Pineda2018}, with only SNe feedback. What \citet{Dobbs2011} and \citet{Tress2020} have in common is that they enforce an artificial SFE $\epsilon$ with observationally motivated values to achieve sensible SFRs. However, why this fairly low SFE exists in the first place is not explained. With our study, we provide physical explanations why the star formation is reduced, namely early feedback from massive stars in form of photo-ionisation and stellar winds.

\citet{Martizzi2016}, \citet{Li2017}, and \citet{Fielding2018} all study stratified disc models, only accounting for SN feedback and without self-gravity. In those simulations, the SNR scales with the SFR inferred from the Kenicutt-Schmidt relation, with one SN exploding per $100-150\,\mathrm{M}_\odot$ of stars. For a gas surface density range comparable to our $\Sigma_\mathrm{gas} = 10 \mathrm{M}_\odot$ between $\Sigma_\mathrm{gas} = 5-30 \mathrm{M}_\odot$ they achieve energy loadings from around $\gamma_\mathrm{sn} \approx 5$ per cent \citep{Martizzi2016}, $\gamma_\mathrm{sn} \approx 15$ per cent \citep{Li2017}, up to $\gamma_\mathrm{sn} \approx 40$ per cent \citep{Fielding2018}. In \citet{Li2017} SNe are distributed evenly in time but randomly in location parallel to the disc. \citet{Fielding2018} seed their SNe clustered in their stratified medium, leading to the creation of super-bubbles and breakouts from the mid-plane ISM. These emphasise the importance of clustered SNe to drive a galactic wind but lack the self-regulation of star formation and are therefore hard to compare with our models. 

The galaxy formation simulations by \citet{Smith2020} model isolated galaxies with a virial mass of $M_\mathrm{vir} = 10^{10}\,\mathrm{M}_\odot$, accounting for SN feedback, photoelectric heating from a spatially varying FUV field and photo-ionisation in HII regions around massive stars with an overlapping Str\"omgren approximation scheme. They find - agreeing with our conclusions - that photo-ionisation has the strongest impact in regulating star formation, insensitive to variations in the star formation threshold density or small-scale efficiency parameters. Furthermore, reducing the clustering of SNe by pre-supernova feedback substantially reduces mass and energy outflow rates. This leads to energy and mass loading factors of $\gamma \sim10^{-3}-10^{-2}$ and $\eta \sim 1-10$, measured at $z = \pm 1\,\mathrm{kpc}$, for the simulations with SN, photoelectric heating and photo-ionisation. Compared to the runs with just SN feedback, the star formation as well as the outflow rates drop by $\sim2$ orders of magnitude when photo-ionisation is added. Photoelectric heating only plays a minor role and does not influence the result to much extent when added to the models with SN-only feedback.

In the \textsc{TIGRESS} simulation suite \citep{Kim2017, Kim2018} star formation is followed self-consistently via a sub-grid sink particle model comparable to our realisation. Their MHD simulations include self-gravity and stellar feedback in form of SNe and photoelectric heating on dust by temporally varying FUV radiation. They assume an optically thin medium with a uniform source distribution in the mid-plane ISM and vary the heating rate temporally, based on the mean radiation that the massive young stars would produce. However, they do not propagate hydrogen ionising UV radiation from massive stars via direct radiative transfer. The assumption of a optically thin medium does not generally hold and the local effects of shielding are not considered, which might result in an over-estimation of the FUV heating. In a setup with $\Sigma_\mathrm{gas} \approx 10\,\mathrm{M}_\odot$, they derive a SFR surface density of $\Sigma_{\dot{M}_\star} \approx 5.13 \times 10^{-3}\,\mathrm{M}_\odot\,\mathrm{yr}^{-1}\,\mathrm{kpc}^{-2}$, lying remarkably close to the observational motivated value of around $\overline{\Sigma}^\mathrm{obs}_{\dot{M}_\star} \approx 4.4 \times 10^{-3}\,\mathrm{M}_\odot\,\mathrm{yr}^{-1}\,\mathrm{kpc}^{-2}$. With this SFR, they achieve moderate mass and energy loading factors of $\eta \approx 1.5$ and $\gamma_\mathrm{sn} \approx 0.05$. Regarding the mass loading, their findings agree with our SN-only model \textit{S}. The star formation properties and energy loading factor, however, are more similar to our \textit{SR} model in star formation ($\Sigma_{\dot{M}_\star} \approx 5.9 \times 10^{-3}\,\mathrm{M}_\odot\,\mathrm{yr}^{-1}\,\mathrm{kpc}^{-2}$), and model \textit{SWR} in energy loading ($\overline{\gamma}_\mathrm{sn} = 0.02$). The possibly boosted FUV heating due the lack of local attenuation might be the reason for the strongly regulated SFR with otherwise only SN feedback. In contrast, we might underestimate the impact of FUV heating by keeping the ISRF at a constant value of $G_0 = 1.7$, independent of the SFR. Another systematic difference compared to our models is the implementation of SNe. For resolved SNe, we inject thermal energy with a fixed radius of 3 grid cells ($\approx 12\,\mathrm{pc}$ in the mid-plane ISM) around the sink particles while leaving the density structure as it is. On the other hand, \citet{Kim2017} set the ambient density of the SN explosions to the mean value of the injection region, in order to prevent over-cooling. We do not find that this difference in the SN implementation explains the difference in the SFR of about one order of magnitude compared to their models to our model \textit{S}. We refer the reader to Appendix \ref{sec:flat} for a short discussion.

The strong effect of ionising UV radiation is also seen in higher resolution simulations on individual cloud scales \citep{Haid2019} and is connected to the finding that, in dense media, ionising UV radiation from massive stars has a stronger impact on the environment than stellar winds \citep{Haid2018}. 
Stellar winds also reduce star formation by limiting star cluster growth \citep{Gatto2017}. However, they do not change the ambient SN densities as dramatically as ionising UV radiation (see Fig. \ref{fig:ambientsn}). We have to note here, that even though we follow a momentum injection scheme for stellar winds the detailed wind bubble structures remain unresolved at the spatial resolution of our simulations. Our conclusions concerning the impact of stellar winds can therefore only be preliminary until higher resolution simulations become available.

Our finding that early feedback in the form of ionising UV radiation and, to a lesser extent, winds is required in order to recover an SFR consistent with the  Kennicutt-Schmidt relation is in good agreement with recent observational results pointing towards the importance of early feedback for regulating star formation. In particular, the short feedback time-scales derived by \citet{Chevance2020} and \citet{Kim2020} for molecular clouds in nearby spiral galaxies are difficult to make consistent with models in which SN feedback dominates but agree well with the predictions of models in which UV radiation play a central role in cloud destruction \citep{Chevance2020b}.

\subsection{Possible caveats}

We fail to fully capture the cold, molecular phase in our simulations. One explanation is that the H$_2$ and CO abundances are likely under-resolved with a ${\Delta x = 3.9\,\mathrm{pc}}$ resolution \citep{Seifried2017, Joshi2019} but also that the molecular gas most likely lives in regions which fulfil the accretion criteria of the sink particles with $n_\mathrm{sink} \approx 10^3$ cm$^{-3}$ and gets absorbed by them \citep[compare with][]{Tress2020}. However, we need to include the chemical network and follow the evolution of the chemical species to properly model the re-processing of the ionising UV radiation.

Galactic shear is not accounted for in our study, as opposed to e.g. \citet{Kim2017, Kim2018}. There are observational studies of molecular clouds in the Milky Way \citep{Dib2012} and of a spiral arm segment of M51 \citep{Schinnerer2017} which do not find strong correlations between star formation activities and shear. On the other hand, theoretical works find a strong impact of differential galactic rotation on the ISM and SFRs \citep{Colling2018}, together with feedback from SNe and HII regions. It is not conclusive how important the inclusion of galactic shear for star formation is. Another effect of large-scale shearing motions is the enhancement the magnetic field strength via small-scale dynamo effects. Earlier studies \citep{Walch2015, Girichidis2016, Pardi2017, Girichidis2018a} show that the main effect of magnetic fields in our setup is the retardation of star formation by counteracting gravitational collapse with magnetic pressure, relatively independent of the magnetic field strength. Also, the anisotropic CR diffusion is insensitive to the strength of the magnetic field but rather its direction \citep{Girichidis2018}. We, therefore, argue that the omission of magnetic field replenishment from a small-scale dynamo driven by large-scale shearing motions in our models does not crucially influence our results.

\section{Summary \& Conclusion} \label{sec:summary}

We present a suite of six stratified galactic disc simulations (Table \ref{tab:runs}), with initial gas surface densities of $\Sigma_{\mathrm{gas}} = 10\,\mathrm{M}_\odot\,\mathrm{pc}^{-2}$, successively accounting for the inclusion of the dominant energy and momentum injection mechanisms of massive stars and all major thermal and non-thermal components of the ISM. The simulations follow 100 Myr of evolution of a turbulently disturbed disk with time-dependent non-equilibrium chemistry, cooling and heating of the dusty, magnetised and self-gravitating ISM, star cluster formation, ionising UV radiation and stellar winds from massive stars, their SN explosions, as well as injection and propagation of CRs. Our study contains the first ISM simulations with self-consistent star (cluster) formation combined with the injection and transport of CRs together with SN feedback and stellar winds and additional ionising UV radiation. Radiative transfer is computed with the novel radiative transfer method \textsc{TreeRay} (W\"unsch et al., submitted) and the N-body dynamics of the sink particle are computed with a recent \nth{4}-order Hermite integrator implementation \citep{Dinnbier2020}. We systematically investigate the impact of the aforementioned stellar feedback processes on star formation, the ISM conditions, and outflow properties. 

The combination of various feedback mechanism from massive stars has non-linear effects on the ISM, star formation, and outflow properties. Only accounting for SN feedback (model \textit{S}) results in an initial starburst $(3.1 \pm 2.2) \times 10^{-2}$ $\mathrm{M}_\odot$ yr$^{-1}$ kpc$^{-2}$ exceeding observed values at similar total gas surface density (Fig. \ref{fig:kennicutt}). The massive star clusters (cluster sink particles with a median mass of $1.6 \times 10^4\,\mathrm{M}_\odot$) form with a high number of massive stars, on average $\overline{N}_\star = 184$ (Table \ref{tab:cluster}). The most massive clusters grow to $10^5\,\mathrm{M_\odot}$, not compatible with estimates for open star cluster masses in the local neighbourhood (Fig. \ref{fig:hist}). This results in strongly clustered SNe with bi-modal ambient density distributions. Early SNe in new clusters explode at high densities ($n_\mathrm{ambient} \approx 30-300\,\mathrm{cm}^{-3}$). The majority of SNe, however, explode at very low ambient densities ($n^{80}_\mathrm{ambient} = 4.2 \times 10^{-3}\,\mathrm{cm}^{-3}$) and generate super-bubbles and a high hot gas volume-filling factor of $\overline{\mathrm{VFF}}_\mathrm{hot} \gtrsim89$ per cent. As a result, the strong outflows deplete the mid-plane ISM rapidly (compare with Fig. \ref{fig:evol_0}, left panel) until star formation ceases towards the end of the simulation (Fig. \ref{fig:sfr}, upper left panel). These outflows are characterised by an average mass loading factor $\eta$ of order unity and an average energy loading factor of $\gamma \gtrsim30$ per cent. 

The inclusion of ionising UV radiation from massive stars has strong consequences for star formation as well as ISM phase structure and outflows even though it does not couple efficiently to the ISM \citep[see][]{Walch2013a, Peters2017, Haid2018}. Ionising UV radiation prevents gas accretion onto cluster sink particles by heating their surrounding ISM, therefore reducing the SFR by about one order of magnitude compared to non-radiation models \citep[see also][]{Peters2017, Butler2017}. For our models, this effect is independent of the inclusion of stellar winds or cosmic rays.
A qualitative comparison of the star cluster masses in our simulations to observational data from \citet{Kharchenko2005} suggests that the accretion limiting effect of ionising UV radiation is needed to achieve star cluster masses comparable to solar neighbourhood conditions (see Fig. \ref{fig:hist}).
Additionally, the ambient SN density distribution becomes uni-modal with most SNe exploding at densities below $\lesssim10^{-1}\,\mathrm{cm}^{-3}$. As a consequence, ionising UV radiation moves all models into the observed regimes for star formation (see Fig. \ref{fig:kennicutt}) and ISM structure of the solar neighbourhood (see Fig. \ref{fig:vffmf}). The formation of HII regions right from the birth of the star clusters decreases the ambient ISM densities of the first SNe. The lower average mass of the cluster as well as the lower number of massive stars per cluster result in reduced mass and energy loading factors of $\overline{\eta} \approx 0.01-0.04$ and $\overline{\gamma}_\mathrm{sn} \approx 0.007-0.013$ (see Table \ref{tab:loadings}). All runs including ionising UV radiation have solar neighbourhood like energy densities for the thermal ($\overline{e}_\mathrm{th} \sim0.6\,\mathrm{erg}\,\mathrm{cm}^{-3}$) and kinetic energy ($\overline{e}_\mathrm{kin} \sim0.4\,\mathrm{erg}\,\mathrm{cm}^{-3}$), with the most complete model \textit{SWRC} being the closest to estimates for the local star-forming ISM from \citet{Draine2011} (see Table \ref{tab:edens}).

In simulations with the strongest outflows like the SN and stellar wind models (\textit{SW, SWC}) CRs have the same effect as reported in more idealised studies before. As soon as the energy outflow is dominated by CRs the outflow changes from being hot gas dominated to warm gas dominated. Simulations including ionising UV radiation have a much lower star formation rate and an outflow driving CR pressure gradient cannot build up during our short simulation time of 100 Myr. However, this changes if the simulations are continued and will be discussed in a follow-up study (Rathjen et al., in prep.) with simulated times up to 300 Myr. CRs have no immediate impact on star formation, stellar cluster properties, or the chemical composition. There is a trend for a $\sim40$ per cent increase in warm gas volume-filling factors, reflecting the trends seen in the outflow. 

Our simulations indicate a qualitative change in the regulation of star formation and the evolution of the star-forming ISM if major stellar feedback processes - in particular the emission of ionising UV radiation - are neglected. If SNe are the only feedback process, star clusters can grow more efficiently leading to a rapid depletion of gas on $\sim100\,\mathrm{Myr}$ time-scales. With mass loading factors of order unity, about the same mass is ejected by outflows and also becomes unavailable for star formation. Both, unhindered star cluster growth as well as galactic outflows regulate the ISM baryon budget and therefore star formation. While the picture of regulating star formation via outflows is generally favoured by cosmological galaxy evolution scenarios \citep[see e.g.][for reviews]{Somerville2015, Naab2017, Tumlinson2017}, it breaks down at the low gas surface densities investigated here when including all major feedback processes of massive stars.

The models including ionising UV radiation not only prevent the initial starburst but favour a different characteristic evolution behaviour in general. At such low surface densities, star formation is entirely controlled by pre-supernova feedback from massive stars on the small-scales of forming star clusters, instead of mid-plane out- and inflows. Mass loading and energy loading factors drop by about one order of magnitude (see Table \ref{tab:loadings}). Our studies, therefore, support previous investigations showing similar trends. 
The realistic model, including stellar winds, ionising UV radiation and CR injection and transport results in the most typical gas phase structure, ISM energy densities, and star formation rates (see e.g. Table \ref{tab:mfvff}, \ref{tab:edens}, \ref{tab:loadings}). Even though the results presented here show clear trends they merely present a status report. Future simulations on longer time-scales, higher resolution and even higher fidelity in physical modelling will have to confirm our conclusions. 

\section*{Acknowledgements}

The authors thank the anonymous referee for the very constructive comments and questions which helped to improve the understanding of the concepts in our models and raised the quality of the manuscript. Furthermore, we gratefully acknowledge the Gauss Centre for Supercomputing e.V. (www.gauss-centre.eu) for funding this project by providing computing time on the GCS Supercomputer SuperMUC-NG at Leibniz Supercomputing Centre (www.lrz.de) under the grant pn34ma. TN acknowledges support from the Deutsche Forschungsgemeinschaft (DFG, German Research Foundation) under Germany's Excellence Strategy - EXC-2094 - 390783311 from the DFG Cluster of Excellence "ORIGINS". PG acknowledges funding from the European Research Council under ERC-CoG grant CRAGSMAN-646955. SW gratefully acknowledges the European Research Council under the European Community's Framework Programme FP8 via the ERC Starting Grant RADFEEDBACK (project number 679852). SW, DS and FD further thank the Deutsche Forschungsgemeinschaft (DFG) for funding through SFB~956 ''The conditions and impact of star formation'' (SW: sub-project C5, and DS: sub-project C6), and SW thanks the Bonn-Cologne-Graduate School. RW acknowledges the support from project 19-15008S of the Czech Science Foundation and from the institutional project RVO:67985815. RSK and SCOG acknowledge financial support from the German Research Foundation (DFG) via the Collaborative Research Center (SFB 881, Project-ID 138713538) ''The Milky Way System'' (subprojects B1, B2, and B8), from the Heidelberg Cluster of Excellence STRUCTURES in the framework of Germany's Excellence Strategy (grant EXC-2181/1 - 390900948) and from the European research Council via the ERC Synergy Grant ECOGAL (grant 855130). The software used in this work was in part developed by the DOE NNSA-ASC OASCR Flash Centre at the University of Chicago \citep{Fryxell2000, Dubey2008}. Visualisations of the simulation results were partly done using the \textsc{yt} library for Python \citep{Turk2011}.

\section*{Data Availability}

The data underlying this article will be available on the \textsc{SILCC} data website at \url{http://silcc.mpa-garching.mpg.de/}, and can be accessed under DR 7.
The derived data underlying this article will be shared on reasonable request to the corresponding author.

\bibliographystyle{mnras}
\bibliography{main}

\appendix

\section{Phase structure of the outflow}\label{sec:phase_outflow}

\begin{figure}
	\centering
	\includegraphics[width=.99\linewidth]{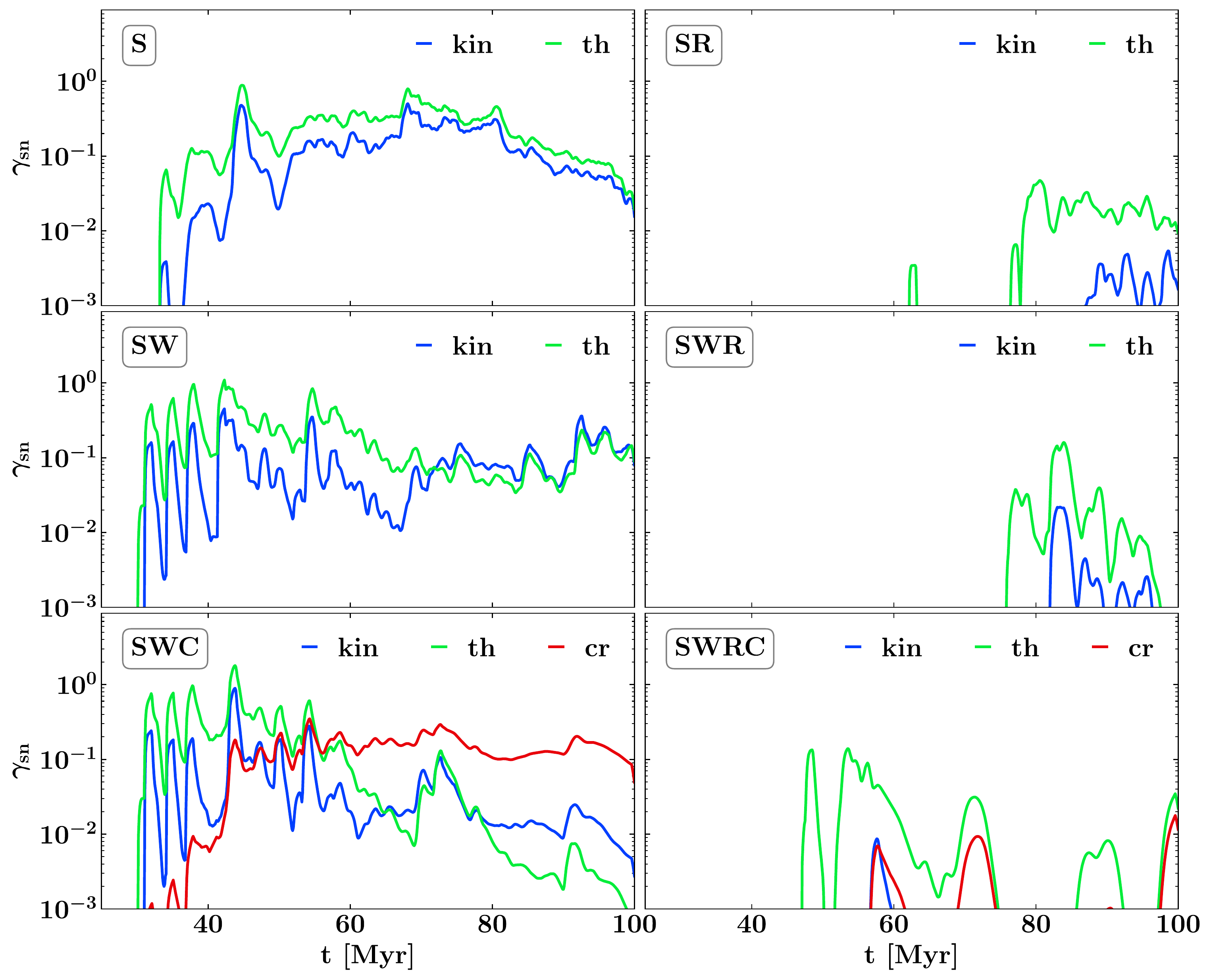}
	\caption{Energy loading factors divided into the energy components kinetic- (blue), thermal- (green), and CR (red) energy vs time for the six models. \label{fig:gamma_i}}
\end{figure}

\begin{figure}
	\centering
	\includegraphics[width=.99\linewidth]{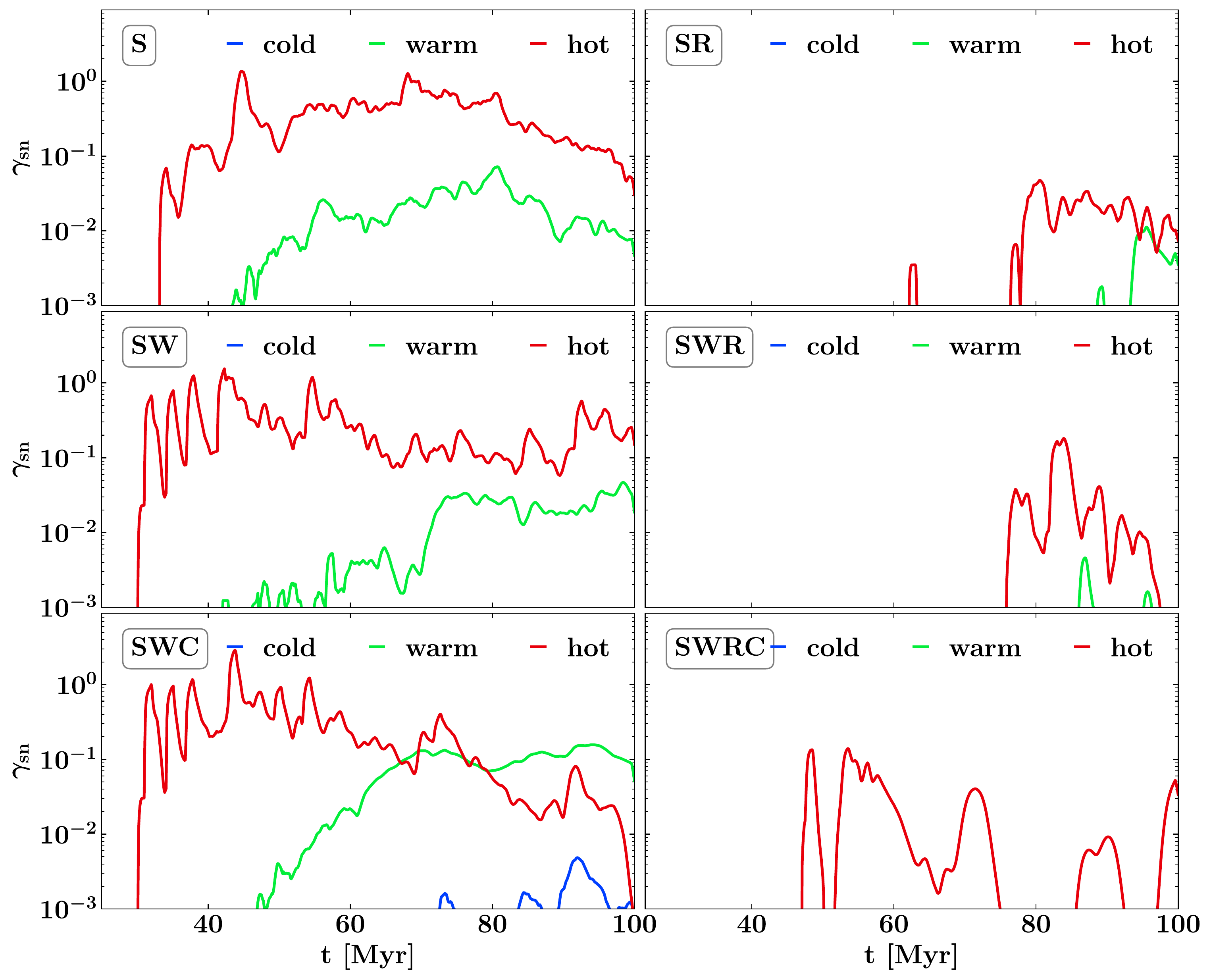}
	\caption{Energy loading factors split up into the contributions of hot- (red), warm- (green), and cold (blue) gas vs time for the six models. \label{fig:gamma_phase}}
\end{figure}

\begin{figure}
	\centering
	\includegraphics[width=.99\linewidth]{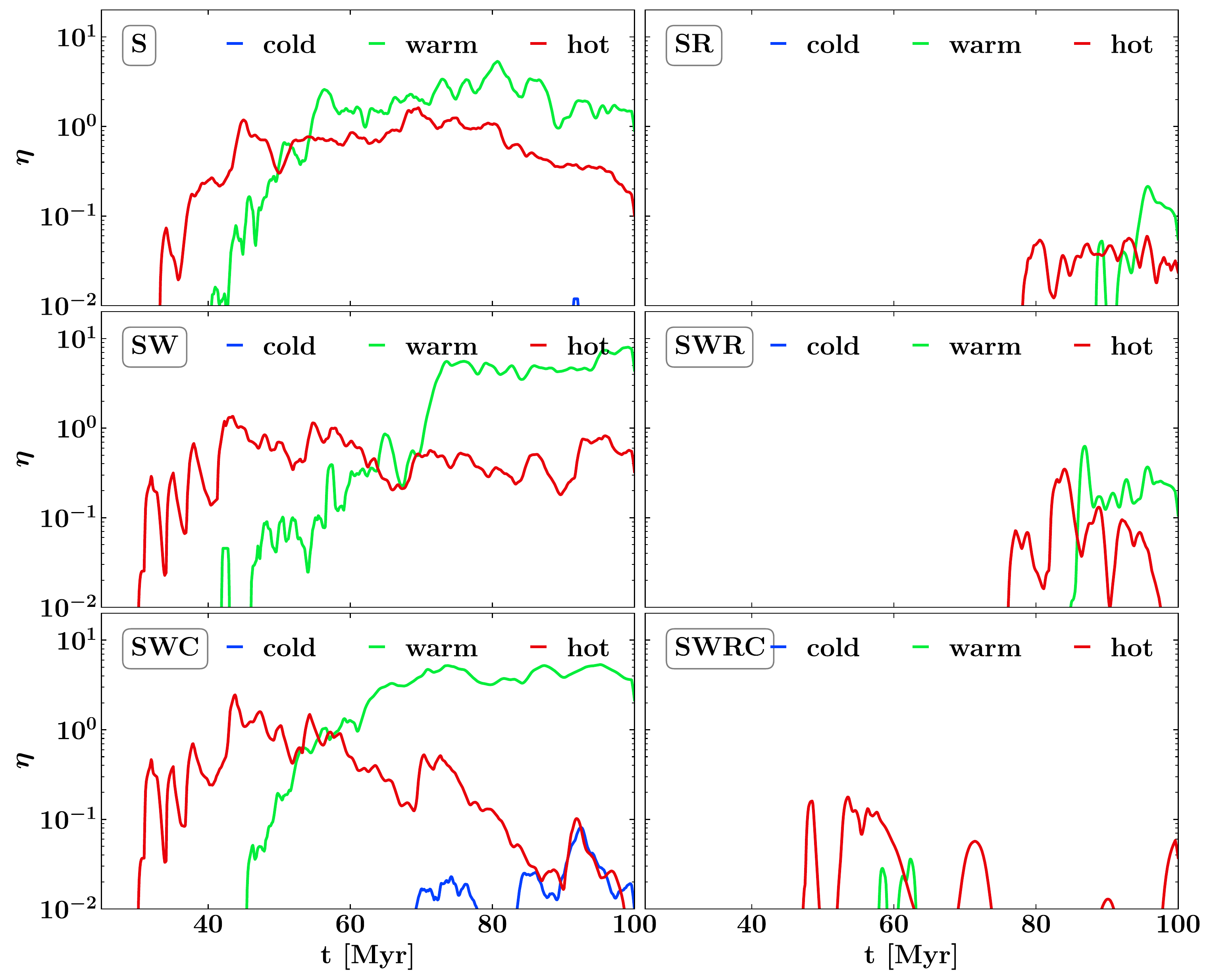}
	\caption{Mass loading factors split up into the contributions of hot- (red), warm- (green), and cold (blue) gas vs time for the six models. \label{fig:eta_phase}}
\end{figure}

We present the time evolutions of the energy-, and mass loading factors divided into the energy components kinetic-, thermal-, and CR energy for $\gamma_\mathrm{sn}$ (Fig. \ref{fig:gamma_i}) and into the contributions of the hot-, warm-, and cold gas phases for $\gamma_\mathrm{sn}$ (Fig. \ref{fig:gamma_phase}) and $\eta$ (Fig. \ref{fig:eta_phase}). The averaged quantities are summarised in Table \ref{tab:loadings}.

The outflows - if present - in the radiation runs (\textit{SR, SWR, SWRC}) are dominated by by thermal energy ($f_{\gamma_\mathrm{th}} \gtrsim 89$ per cent. For the runs without radiation and higher SFR (\textit{S, SW}) the ratio between thermal and kinetic energy in the outflow is $\sim2:1$. The CR run without radiation (\textit{SWC}) starts with a similar ratio but the outflow quickly becomes governed by the CR energy. Due to the lower SFR in \textit{SWRC}, and hence lower CR injection rate, the raise in CR energy in the outflow is only seen at the very end of the simulation (Fig. \ref{fig:gamma_i}).

The thermal composition of the energy outflow (Fig. \ref{fig:gamma_phase}) is initially dominated by the fast moving hot gas phase, with the slower moving warm phase gas catching up with some time delay. The same is also true for the mass outflows (Fig. \ref{fig:eta_phase}). A cold gas outflow with a mass loading factor between $\eta \sim 1-10$ per cent is only present at later stages in \textit{SWC}, supported by the additional CR pressure gradient. We expect - like it is demonstrated in idealised studies \citep{Girichidis2016, Girichidis2018} - that a significant cold, and possibly even molecular, gas outflow will develop later in \textit{SWRC}, when more star formation has happened. This will be studied in Rathjen, et al. (in prep.).

\section{Energy injection}\label{sec:inj}

\begin{figure}
	\centering
	\includegraphics[width=.99\linewidth]{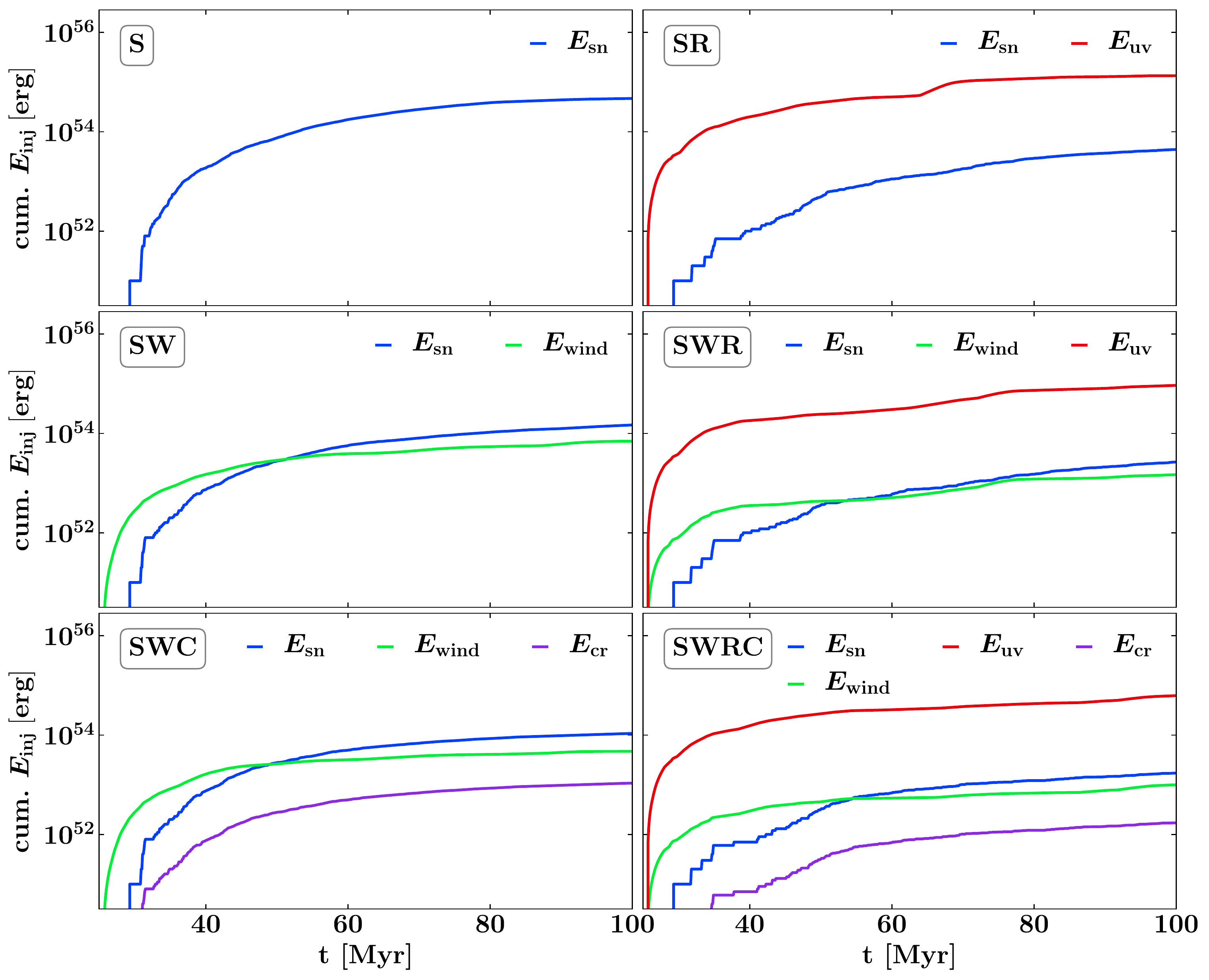}
	\caption{Cumulative energy injection over time of the different stellar feedback mechanisms - SNe, winds, UV radiation and CR - in our six models. \label{fig:inj}}
\end{figure}

For completeness, we show the cumulative injected energy $\dot{E}_\mathrm{inj}$ in our six models as a function of time in Fig. \ref{fig:inj}. As discussed in Sec. \ref{sec:gamma}, the injected energy by stellar winds and SNe are of the same order of magnitude. The wind injection is continuous throughout a massive star's lifetime, while SN injection is instantaneous at the end of a massive star's lifetime. The total injected CR energy is 10 per cent of the SN energy by construction. The total energy budget is dominated by the UV photons luminosity by up to 2 orders of magnitude in radiation runs \textit{SR, SWR, SWRC}.

\section{Density flattening}\label{sec:flat}

\begin{figure}
	\centering
	\includegraphics[width=.99\linewidth]{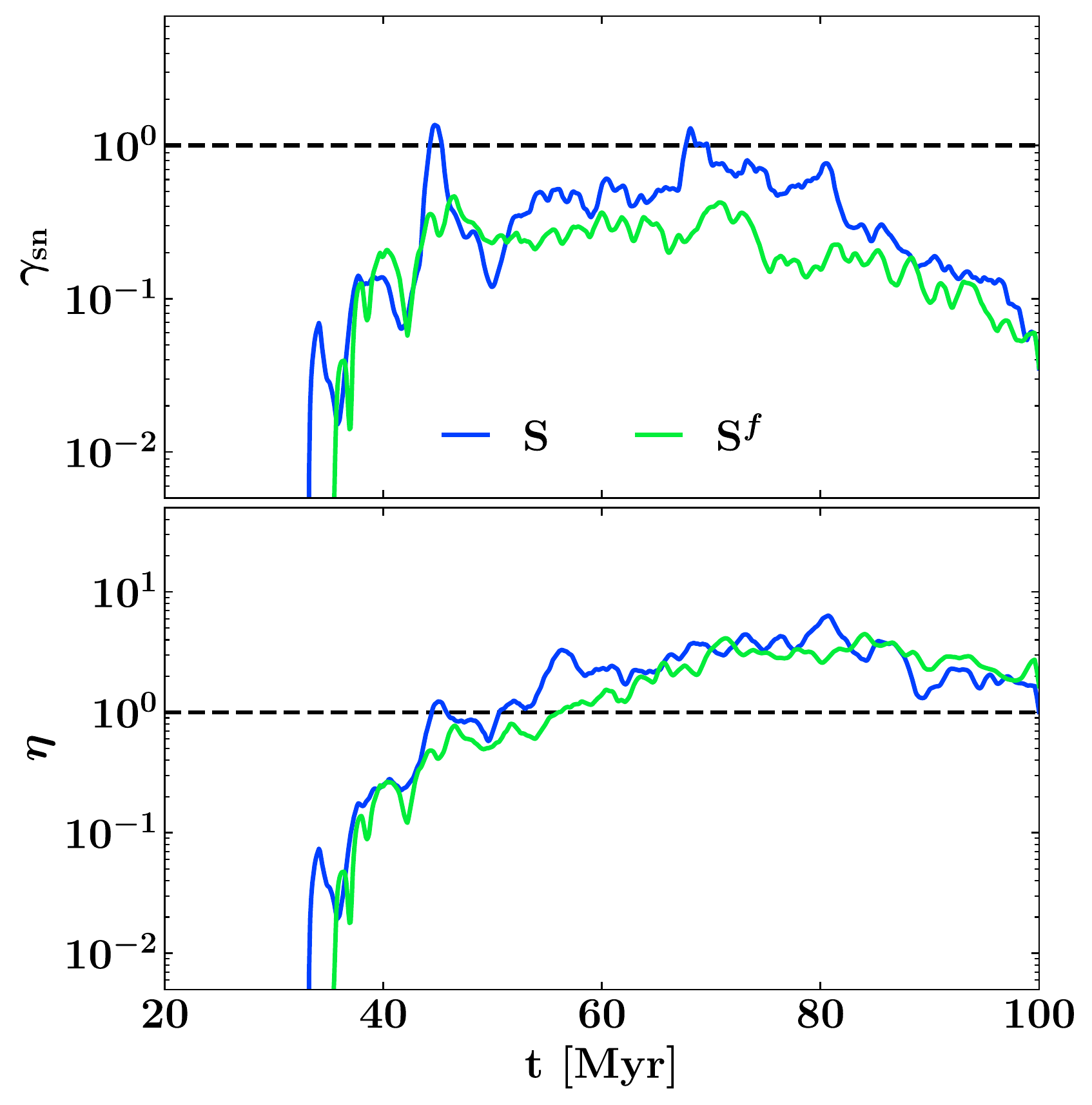}
	\caption{Energy- and mass loading factors over time for the two SN-only runs with different approaches of injecting the thermal SN energy into the surrounding medium - with (S$^f$) and without (S) density flattening of the injection region prior the injection. \label{fig:flatten}}
\end{figure}

We realise SNe with resolved Sedov-Taylor phase by injecting thermal energy ($10^{51}\,\mathrm{erg}$) in a spherical region with radius of 3 grid cells around the sink particle in which a SN explodes, as described in Sec. \ref{sec:methods}. The density and temperature structure around the particle at the moment of injection is kept intact. However, another possibility is to \textit{flatten} the injection region by setting the mass density, momentum density, and internal energy in that region to their mean values before assigning additional momentum and thermal energy like it is done in \citet{Kim2017, Kim2018}. They argue that this step is needed for self-regulation of the SFR and to prevent over-cooling.
We test this with a model \textit{S}$^f$, in which we only include SN feedback like in \textit{S} but set the ambient gas density of the injection region around a SN to the mean value of this region. This results in a reduced SFR surface density by a factor of $\sim1.5$, a reduced SN energy loading factor by a factor $\sim2$ but nearly identical mass loading factors. The average number of massive stars per cluster is reduced by a factor of $\sim1.6$ from 184 in \textit{S} to 115 in \textit{S}$^f$ (see Table \ref{tab:flat} and Fig. \ref{fig:flatten}).
Those results suggest that \textit{density flattening} is not needed to prevent over-cooling when realising SNe with a fixed injection radius. With density flattening and no other feedback processes but SNe at play, we still achieve a fairly high SFR surface density of $\sim2\times10^{-2}\,\mathrm{M}_\odot\,\mathrm{yr}^{-1}\,\mathrm{kpc}^{-2}$, which lies in the upper limit regime of observed SFR surface densities for gas surface densities around $10\,\mathrm{M}_\odot\,\mathrm{pc}^{-2}$ (compare with Fig. \ref{fig:kennicutt}). 

\begin{table}
    \caption{Average SFR surface density, SN energy loading factors, mass loading factor, number of massive stars per cluster for model \textit{S} and a run with all the same parameters but the realisation of density flatting in the SN injection region \textit{S}$^f$, each with 1$\sigma$.}
    \begin{tabular}{lcccc}
        \hline
        Run & $\overline{\Sigma}_{\dot{M}_\star}$ & $\overline{\gamma}_\mathrm{sn}$ & $\overline{\eta}$ & $\overline{N_\star}$ \\
         & [$\mathrm{M}_\odot$ yr$^{-1}$ kpc$^{-2}$] & [\%] & & \\
        \hline
        \textit{S}     & $(3.1 \pm 2.2) \times 10^{-2}$ & $33.6 \pm 30.3$ & $1.94 \pm 1.51$ & $184$ \\
        \textit{S}$^f$ & $(2.0 \pm 1.1) \times 10^{-2}$ & $18.1 \pm 12.2$ & $1.65 \pm 1.33$ & $115$\\
        \hline
    \end{tabular}
    \label{tab:flat}
\end{table}
        
\bsp
\label{lastpage}
\end{document}